\documentclass[12pt]{article}
\usepackage{fullpage,authblk}
\usepackage{amsmath,amsthm,graphicx,amssymb,verbatim,listings}
\usepackage{wasysym,color,enumitem,mathcomp}
\usepackage[T1]{fontenc}     
\usepackage{lmodern, tipa}   
\usepackage[ pdftex, plainpages = false, pdfpagelabels,
                 pdfpagelayout = useoutlines,
                 bookmarks,
                 bookmarksopen = true,
                 bookmarksnumbered = true,
                 breaklinks = true,
                 linktocpage=all,
                 pagebackref=false,
                 colorlinks = true,
                 linkcolor = Blue,
                 urlcolor  = blue,
                 citecolor = BrickRed,
                 anchorcolor = green,
                 hyperindex = true,
                 hyperfigures
                 ]{hyperref}
\usepackage[usenames, dvipsnames]{xcolor}
\usepackage{xifthen} 

\newcommand{\tr}{\hbox{tr}}
\newcommand{\moment}[1]{{\ensuremath{\left\langle #1 \right\rangle}}}
\newcommand{\ket}[1]{{\ensuremath{\left| #1 \right\rangle}}}
\newcommand{\bra}[1]{{\ensuremath{\left\langle #1 \right|}}}
\newcommand{\braket}[2]{{\ensuremath{\left\langle #1 \middle| #2
      \right\rangle}}}

\newcommand{\arxiv}[2][]{\ifthenelse{\isempty{#1}}{\href{http://arxiv.org/abs/#2}{{\tt arXiv:\allowbreak{}#2}}} {\href{http://arxiv.org/abs/#2}{{\tt arXiv:\allowbreak{}#2 [#1]}}}}
\newcommand{\pirsa}[1]{\href{http://pirsa.org/#1/}{{\tt PIRSA:\allowbreak{}#1}}}
\newcommand{\booktitle}{\textsl}
\newcommand{\hrefdoi}[2]{\href{https://dx.doi.org/#1}{#2}}

\begin{document}
\title{FAQBism}
\author[$\dag\star$]{John B.\ DeBrota}
\author[$\dag$]{Blake C.\ Stacey}
\affil[$\dag$]{\small{Department of Physics, University of Massachusetts Boston, 100 Morrissey Boulevard, Boston MA 02125, USA}}
\affil[$\star$]{Stellenbosch Institute for Advanced Study (STIAS), Wallenberg Research Center at Stellenbosch University, Marais Street, Stellenbosch 7600, South Africa}

\date{\small\today}

\maketitle

\abstract{We answer several questions that have been Frequently Asked about QBism. These remarks (many of them lighthearted) should be considered supplements to more systematic treatments by the authors and others.}

\bigskip

\begin{description}
    \item \textborn\textbf{QBism}
      \textbackslash\textipa{\textprimstress}ky\"u-\textipa{\textsecstress}bi-z\textipa{@}m\textbackslash
      \ \textit{n}. an interpretation of quantum mechanics in which
      the ideas of \emph{agent} and \emph{experience} are
      fundamental. A ``quantum measurement'' is an act that an agent
      performs on the external world. A ``quantum state'' is an
      agent's encoding of her own personal expectations for what she
      might experience as a consequence of her actions. Moreover, each
      measurement outcome is a personal event, an experience specific
      to the agent who incites it. Subjective judgments thus comprise much of the quantum machinery, but the formalism of the theory establishes the standard to which agents should strive to hold their expectations, and that standard for the relations among beliefs is as objective as any other physical theory.
      \end{description}
\bigskip
\pagebreak
As proponents of an interpretation of quantum mechanics, we are
accustomed to encountering puzzlement. We know that this is an
unusual, perhaps even disreputable activity for professional
physicists to engage in. One of the authors came to QBism from quantum
chemistry, and the other from nonequilibrium statistical physics. We
are here largely because we sought to apply the same habits to quantum
foundations that we would find virtuous in any other field of
physics:\ separating principles from convenient conventions,
reformulating old mathematics in new ways to make a different set of
questions easy, maintaining a healthy disregard for philosophers'
judgments about what is impossible. Yet even a mostly ordinary
upbringing can lead to a surprising place.

What follows is our attempt to take the questions we have encountered,
select those that have been posed in good faith, and provide responses
that summarize QBist thinking on those topics.  We have tried to be
serious without being dour.  Finding quantum foundations important at
all may be controversial among physicists, potentially more so than
any specific choice of interpretation.  More shocking still is the
suggestion that the material be approached with a sense of fun and
adventure.

\begin{enumerate}[label=\textbf{\color{Blue}\theenumi.}]
    \item \hyperref[sec:name]{\textbf{Why did you change from Quantum Bayesianism to QB\lowercase{ism}?}}\\
        \textit{There are too many different kinds of ``Bayesian''.}
    \item \hyperref[sec:local]{\textbf{Why do you call QB\lowercase{ism} ``local''?}}\\
        \textit{That which happens, happens locally; that which changes ``spookily'' is not physical.}
    \item \hyperref[sec:superposition]{\textbf{What does it feel like to be in a quantum superposition?}}\\
    \textit{Nothing --- wave functions are my personal expectations for the consequences of my freely chosen actions.}
    \item \hyperref[sec:solipsism]{\textbf{Isn't it just solipsism?}}\\
    \textit{No. The existence of an external world is a central postulate of QBism.}
    \item \hyperref[sec:copenhagen]{\textbf{Isn't it just the Copenhagen interpretation?}}\\
    \textit{No. There are considerable differences between the views
      of each of the founders and QBism.}
    \item \hyperref[sec:interpret-prob]{\textbf{Why does the
        interpretation of probability theory matter?}}\\
      \textit{Although it's the kind of question physicists like to
        gloss over, it's an inextricable part of making progress.}
    \item \hyperref[sec:doubleslit]{\textbf{ What is the meaning of the double-slit experiment in QBism?}}\\
        \textit{It's a canonical, helpful example of probabilities meshing together in a nonclassical way, but it isn't the deepest such.}
\item \hyperref[sec:pbr]{\textbf{ Doesn't the PBR theorem prove QBism wrong?}}\\
        \textit{No. It just doesn't apply to QBism, and even its inventors
          don't think that it does.}
    \item \hyperref[sec:bayesrule]{\textbf{ Is QBism about the Bayes rule?}}\\
      \textit{No. In QBism, being ``Bayesian'' is about something more fundamental.}
    \item \hyperref[sec:technical]{\textbf{What technical questions have been motivated by QBism?}}\\
      \textit{Examples include quantum de Finetti theorems for ``unknown states'' and ``unknown processes'' as well as compatibility criteria for quantum states. Most recently, QBism has prompted work on Symmetric Informationally Complete quantum measurements (SICs). The lessons they teach about how probability works in quantum theory keep growing more interesting.}

    \item \hyperref[sec:noncommute]{\textbf{Isn't quantum probability just classical probability but
        noncommutative?}}\\
      \textit{A careful analysis reveals that one has to dig deeper
        than that.}
    \item \hyperref[sec:decoherence]{\textbf{Doesn't decoherence solve
        quantum foundations?}}\\
      \textit{Momentarily deferring a question is not the same as answering it.}
    \item \hyperref[sec:rqm]{\textbf{Is QBism like Rovelli's ``Relational Quantum Mechanics''?}}\\
        \textit{In a few ways, yes, but there are important differences and it's hard to pin down RQM on some significant points.}
    \item \hyperref[sec:cox]{\textbf{Why do QBists prefer de Finetti over Cox?}}\\
        \textit{The Cox approach is too loaded in the direction of thinking of inferences regarding hidden variables.}
    \item \hyperref[sec:finite]{\textbf{Why so much emphasis on finite-dimensional Hilbert spaces?}}\\
        \textit{It's a good place to look for the essential quantum mysteries.}
    \item \hyperref[sec:objprob]{\textbf{Aren't the probabilities in
        quantum physics objective?}}\\ \textit{Ultimately,
      probabilities just can't work that way. But that's fine, because
      physicists don't need them to.}
    \item \hyperref[sec:agent]{\textbf{Don't you have to define ``agent''?}}\\
        \textit{No, for the same reason you don't have to in any standard decision-theoretic situation.}
    \item \hyperref[sec:future]{\textbf{Where and when does the agent end and the external world begin?}}\\
        \textit{What matters is that you are consistent about the notion of an agent within each scenario you try to reason about.}
    \item \hyperref[sec:explain]{\textbf{Does QBism lose the ``explanatory power'' of other interpretations?}}\\
        \textit{Not with respect to a reasonable notion of ``explanation''.}
    \item \hyperref[sec:insufficient]{\textbf{Aren't probabilities an insufficient representation of beliefs?}}\\
        \textit{They might not be sufficient in the final analysis,
          but we don't need them to be.}
    \item \hyperref[sec:unitary]{\textbf{What does unitary time
        evolution mean in QBism?}}\\ \textit{Mathematically and
      conceptually, unitaries are conditional probabilities, used in a
      way that respects the Born Rule.}
    \item \hyperref[sec:wigner]{\textbf{What do the recent Extended Wigner's
        Friend thought-experiments imply for QBism?}}\\
      \textit{So far, the mini-field of wrapping a Wigner's Friend
        scenario around a Bell-type test is still too undeveloped to
        go beyond what QBism has already learned from Wigner and Bell.}
    \item \hyperref[sec:mwi]{\textbf{Is QBism compatible with the Many
        Worlds Interpretation?}}\\ \textit{No, our view and those of
      the Everettian creeds are genuinely contradictory.}
\item \hyperref[sec:others]{\textbf{What are good things to read about QBism?}}\\
        \textit{The Stanford Encyclopedia of Philosophy, von Baeyer's book, and others.}
\end{enumerate}

\section{Why did you change from Quantum Bayesianism to QB\lowercase{ism}?}
\label{sec:name}
Originally, the \emph{Q} stood for \emph{Quantum} and the \emph{B} for
\emph{Bayesian.} The former is still true.

Back in the 1990s and early 2000s, the term ``Quantum Bayesianism''
was serviceable.  However, it had its issues.  For one thing, nobody
was consistent on whether to capitalize the \emph{Q}: Those who called
themselves Quantum Bayesians preferred it uppercase, so that neither
half of the term had undue emphasis over the other, but try to
convince a copy editor of that simple point!  More importantly, there
are many varieties of Bayesianism, and plenty of self-declared
Bayesians disagreed in fundamental ways with the particular variety
that our school found necessary for quantum physics.  For a while,
N.\ David Mermin joked that the \emph{B} should stand for \emph{Bruno}
de Finetti~\cite{Mermin:2013}, and Chris Fuchs suggested that the
\emph{QB} was like \emph{KFC,} which once stood for ``Kentucky Fried
Chicken'' but is now a stand-alone trademark~\cite{Gefter2015}.

More recently, we found a way to expand the \emph{B} that we had never
anticipated --- a rolling, Lewis-Carroll-esque word:
\emph{bettabilitarianism}~\cite{Fuchs:2016}.  This word comes, of all
places, from the jurist Oliver Wendell Holmes, Jr. To quote Louis Menand's
history of American pragmatism, \booktitle{The Metaphysical Club:\ A
  Story of Ideas in America}:
\begin{quote}
`The loss of certainty' is a phrase many intellectual historians have used to characterize the period in which Holmes lived. But the phrase has it backward. It was not the loss of certainty that stimulated the late-nineteenth-century thinkers with whom Holmes associated; it was the discovery of uncertainty. Holmes was, in many respects, a materialist. He believed, as he put it, that ``the law of the grub \ldots\ is also the law for man.'' But he was not entirely a determinist, because he did not think that the course of human events was fixed \ldots. Complete certainty was an illusion; of that he was certain. There were only greater and lesser degrees of certainty, and that was enough.  It was, in fact, better than enough; for although we always want to reduce the degree of uncertainty in our lives, we never want it to disappear entirely, since uncertainty is what puts the play in the joints.  Imprecision, the sportiveness, as it were, of the quantum, is what makes life interesting and change possible. Holmes liked to call himself a ``bettabilitarian'':  we cannot know what consequences the universe will attach to our choices, but we can bet on them, and we do it every day.
\end{quote}
A QBist declares, ``I strive to be the very model of a Quantum
Bettabilitarian!''

We have occasionally seen manglings like ``QBian'' and even
``Qubian''. These spoil the pun of \emph{QBism} and are thus strongly
deprecated.

\section{Why do you call QB\lowercase{ism} ``local''?}
\label{sec:local}
A journey of a thousand perspective shifts begins with a single step.
In this case, the first step is to realize why the scenarios trotted
out to imply ``quantum nonlocality'' actually don't~\cite{Fuchs:2013}.
The standard argument for ``nonlocality'' rests upon entanglement.
Conjure up a pair of qubits, for example, assign to them a maximally
entangled state and ship one of the pair off to Mars.  Upon measuring
the qubit left on Earth, ``the state of the qubit on Mars changes
instantaneously''.  But this is not a \emph{physical} change of any
property of a material object!  Compare this with classical
electromagnetism: In that subject, if we could toggle a quantity at a
distance but only in ways that could not effect a transmission of
information, we'd have no hesitation in calling that quantity
unphysical --- an artifact, we'd say, of choosing a gauge that does
not respect relativistic causality.  QBism says that the right way to
interpret quantum theory is to take this helpful and uncontroversial
move seriously.  When Alice measures her Earth-bound qubit, what
changes for its partner on the red planet?  Only \emph{Alice's
  expectations} for what might happen \emph{to Alice} herself,
\emph{if} she were to make the journey and intervene upon that qubit.

The fact that nature violates Bell inequalities is reason to reject
the hypothesis used to derive those inequalities, ``local
realism''. But when we decide to adopt \emph{non-(local realism)}, we
have a choice of how to clear those parentheses.  We can put the
\emph{non-} on either half, and when we consider the highly
specialized character of what is actually meant by \emph{realism} in
this context, keeping the \emph{local} turns out to be the natural
move. Another way of saying this is that adopting quantum theory does
not force us to revise the notions of ``causal structure'' developed
in classical physics, such as the conceptual tool of Minkowski
spacetime.

\section{What does it feel like to be in a quantum superposition?}
\label{sec:superposition}
A QBist affirms, ``My quantum states are mine, your quantum states are yours. If someone else considers a quantum system containing me and ascribes to that system a quantum superposition state, so be it. That is their quantum state assignment. It doesn't make sense for me to assign myself a quantum state if a quantum state is an encoding of my own beliefs for the outcomes of my freely chosen actions.'' If something feels off about this answer, consider whether you are assuming that there is a
``correct'' --- i.e., purely physically mandated --- quantum state in this scenario. For a QBist, there is never such a quantum state just as in personalist probability theory there is never an ontologically ``correct'' probability distribution. The answer to the question ``What does it feel like to be in a quantum superposition?''\ is the same as the answer to the question ``What does it feel like to be in someone else's probability distribution about me?''.

Incidentally, this is one problem with the no-go argument made by
Frauchiger and Renner, originally intended to rule out what they
called ``single-world interpretations'' of quantum
theory~\cite{Frauchiger:2017}.  This argument, as well as others
closely related to it~\cite{Brukner:2017}, all at some point make an
assumption that amounts to an agent putting herself into a quantum
superposition.  Ultimately, this makes no more sense than trying to
crawl inside a probability distribution and live there.

\section{Isn't it just solipsism?}
\label{sec:solipsism}
On the issue of solipsism, QBism stands with Martin
Gardner~\cite{Gardner83}:
\begin{quotation}
\noindent The hypothesis that there is an external world, not
dependent on human minds, made of {\it something,} is so obviously
useful and so strongly confirmed by experience down through the ages
that we can say without exaggerating that it is better confirmed than
any other empirical hypothesis.  So useful is the posit that it is
almost impossible for anyone except a madman or a professional
metaphysician to comprehend a reason for doubting it.
\end{quotation}
For a QBist, the basic subject matter of quantum theory is \emph{an
  agent's interactions with the outside world}; the formalism of
quantum theory makes no sense otherwise.  Were there no systems
outside the QBist's mind, there would be no interface between agent
and world, and quantum theory would have no subject matter.  QBism is
a full-throated rejection of metaphysical solipsism.

Someone once asked us, ``Where is the real world in such a view?'' The
real world is exactly where it always has been. It is the world in
which our species evolved. It is the world in which we grow and strive
and protest, where we learn by individual experience --- including our
encounters with the words of others --- the pain of heartbreak and the
utility of the Lorentz transform. It has conditioned our calculus of
expectations, even as those expectations themselves remain intensely
personal.

To a QBist, ``measurement'' is that variety of interaction which
physics understands best, precisely because experiments are
actions whose potential outcomes we can catalogue.
Measurements are to quantum physics what ``model organisms''
are to biology.  Why do developmental biologists know so much
about zebrafish?  Because their embryos are transparent!  But
life flourishes on, unanalysed, beyond our microscopes.  Far from
being solipsistic, QBism recognises just how little of nature we
have managed to touch.

It is true that QBists refuse to make an upfront definite claim about
what the stuff of the world is. How then, can they have a consistent
doxastic interpretation?\footnote{A brief note on useful terminology:
  \emph{epistemic} refers to knowledge and information,
  \emph{doxastic} to belief, and \emph{ontic} to brute elements of
  physical reality.} This is accomplished by being clear on what
quantum information is actually information about: A quantum state
encodes a user's beliefs about the experience they will have as a
result of taking an action on an external part of the world. Among
several reasons that such a position is defensible is the fact that
any quantum state, pure or mixed, is equivalent to a probability
distribution over the outcomes of an informationally complete
measurement~\cite{DeBrota18}. Accordingly, QBists say that a quantum
state is conceptually no more than a probability distribution. Okay,
fine, but what is the stuff of the world? QBism is so far mostly
silent on this issue, but \emph{not} because there is no stuff of the
world. The character of the stuff is simply not yet understood well
enough. Answering this question is the goal, rather than the premise.

Is this an unacceptable weakness of the interpretation? Well,
that's a matter of opinion, but ours is that it is not. Must
we demand that a complete ontology be laid out before one's ramblings
graduate to the status of an ``interpretation''? If taken to the
extreme, this is clearly unfair: One might claim that no one has a
qualifying interpretation because we don't have a successful theory
for quantum gravity and so every proposed ontology \emph{necessarily}
fails. More practically, feeling pressured to commit to an ontology
prematurely may leave physicists unable to imagine one which departs
sufficiently from classical intuitions. Why not see if the right
ontology can be teased out from the formalism itself and a principled
stance on the meaning of its more familiar components (such as
probability distributions)?

In fact, QBism has had ontological aspirations ever since the
beginning.  (It's hard to have ontological aspirations for a theory if
you think you'd have to be a solipsist to hold it.)  There are
\emph{structural realist} and \emph{neutral pluralist} elements in
QBism, and there seems to be a process or event ontology underlying it
all, somewhere in a spectrum of things suggested by William James,
Henri Bergson, Alfred North Whitehead, and John Archibald
Wheeler~\cite{Atmanspacher:2014}.  \emph{The stuff of the world is the
  becomings of the world.}  However, we really don't believe we'll be
able to say anything in proper detail until we get the quantum
formalism into a better shape.  (That's what all the SIC research
described in~\S\ref{sec:technical} is about.)  So, from this
perspective, QBism is a project.

Mermin \cite{Mermin14} argues,
\begin{quotation}
\noindent QBists are often charged with solipsism: a belief that the
world exists only in the mind of a single agent.  This is wrong.
Although I cannot enter your mind to experience your own private
perceptions, you can affect my perceptions through language.  When I
converse with you or read your books and articles in
\booktitle{Nature}, I plausibly conclude that you are a perceiving
being rather like myself, and infer features of your experience.  This
is how we can arrive at a common understanding of our external worlds,
in spite of the privacy of our individual experiences.
\end{quotation}

This leads us to an important topic: communication between agents.
What does the idea of agents comparing notes mean when we interpret
quantum mechanics as a single-user theory?  Consider an agent Alice.
She can use quantum mechanics as a ``manual for good living'', a way
to organize her expectations while navigating an irreducibly
unpredictable world.  Alice encounters a system which she designates
as ``Bob''.  Alice can ask Bob about his experiences and use quantum
mechanics to predict his answer.  If Alice so chooses, she can
incorporate Bob's responses into her expectations.  Nothing in the
formalism of quantum theory forces her to do so, however.

\section{Isn't it just the Copenhagen intepretation?}
\label{sec:copenhagen}
This has been said a lot through the years, and we continue to hear it
today.  Sometimes, it's said that QBism is trying to be more
Copenhagen than the Copenhagen interpretation itself. As if QBism had
a fever, and the only prescription were more Copenhagen! But the idea
that there ever was a unified ``Copenhagen interpretation'' --- i.e.,
that the definite article is remotely applicable --- was a myth of the
1950s. Trying to exceed ``the'' Copenhagen interpretation in any
respect is to race against a phantom.

QBism does not have, for example, Bohr's emphasis on ``ordinary
language''~\cite{Stacey:2018b}, whatever that might mean. Nor does it
have the quantum-classical cut of Heisenberg, the classical laboratory
equipment of Landau and Lifshitz~\cite{Mermin:2018}, the public
experimental records of Pauli~\cite{notwithstanding}, the essentially
ontic state vectors of early Bohm~\cite{Bohm:1951}, or the frequentism
of early von Neumann~\cite{Stacey:2016b}. Unlike van Kampen, QBism
does not presume that the vanishing of interference terms will solve
all riddles~\cite{VanKampen:1988}. Unlike Wheeler, QBism does not
posit that all observers should ideally have the same information
about a system and thus the same quantum state for
it~\cite[footnote 9]{notwithstanding}. There simply is not a way to
summarize this overflow of differences by claiming that QBism is
``more Copenhagen''.

At one point, the Wikipedia article on QBism claimed that it ``is very
similar to the Copenhagen interpretation that is commonly taught in
textbooks''. What does this even mean? First, as we noted, there's no
such thing as ``the'' Copenhagen interpretation. In addition, claiming
that ``the Copenhagen interpretation'' is ``commonly taught in
textbooks'' conflates the early developers of quantum theory
\emph{and} the varied modern expositions of it into a vague
mishmash. Asher Peres' textbook is more instrumentalist than the
undergraduate standards; the \booktitle{Feynman Lectures} handle
probability in a less frequentist way than Peres. Are all common
textbooks Copenhagen, or is Copenhagen that which is commonly taught
in all textbooks? Better to strike the term ``Copenhagen
interpretation'' from our lexicon going forward and instead be precise
about what views we mean!

\section{Why does the interpretation of probability theory matter?}
\label{sec:interpret-prob}
E.\ T.\ Jaynes put the basic point rather well~\cite{Jaynes:1990}:
\begin{quotation}
  \noindent [O]ur present QM formalism is not purely epistemological;
  it is a peculiar mixture describing in part realities of Nature, in
  part incomplete human information about Nature --- all scrambled up
  by Heisenberg and Bohr into an omelette that nobody has seen how to
  unscramble. Yet we think that the unscrambling is a prerequisite for
  any further advance in basic physical theory. For, if we cannot
  separate the subjective and objective aspects of the formalism, we
  cannot know what we are talking about; it is just that simple.
\end{quotation}
According to Jaynes, the way to unscramble that Heisenberg--Bohr
omelette will be ``to find a different formalism, isomorphic in some
sense but based on different variables''~\cite{Jaynes:1989}.

The \emph{results} of quantum-mechanical calculations are generally
probabilities, or stand-ins for them like rates and effective
cross-sections, and so the question ``what means probability?''\ must
be addressed sooner or later. This question gains urgency when we
realize that the \emph{inputs} to those calculations are just as
probabilistic as the outputs. We can see this concretely by focusing
on the simplest possible quantum system, a single qubit.  A quantum
state for a qubit can be written as a linear combination of the Pauli
matrices, where the coefficients are expectation values for the
outcomes of the three Pauli measurements:
\begin{equation}
  \rho = \frac{1}{2}\left(
  I + \moment{x} \sigma_x + \moment{y}\sigma_y + \moment{z}\sigma_z
  \right).
\end{equation}
Because each Pauli measurement has only two possible outcomes, $+$ and
$-$, we can write the expectation value $\moment{x}$ as
\begin{equation}
  \moment{x} = p(+|x) - p(-|x) = 2p(+|x) - 1,
\end{equation}
and similarly for $\moment{y}$ and $\moment{z}$. Mathematically, any
qubit state $\rho$, whether pure or mixed, is nothing more than a
convenient packaging of the three probabilities $p(+|x)$, $p(+|y)$ and
$p(+|z)$.  Thus, whatever status one grants to quantum states, one
must grant that same status to at least some
probabilities. Conversely, if a particular interpretation of
probability theory turns out to be logically untenable, then that
rules out a possible way of interpreting quantum states, too.

Note that the fact that specifying a qubit quantum state requires
\emph{three} probability values is more fundamental than any choice of
those values. To put it another way, when we take our empirically
successful theory of physics and find the simplest case where it
applies, we see that the theory has three knobs, not one or two or
five. This is a more primitive, more basal statement about our theory
than a choice of $\rho$ is!

Modern quantum information theory provides an even deeper take: Any
quantum state can in fact be specified, not just as a compendium of
probabilities for different experiments, but as a probability
distribution over the outcomes of a \emph{single} experiment. For more
on the concept of a \emph{reference measurement}, see
\S\ref{sec:technical}.

\section{What is the meaning of the double-slit experiment in QBism?}
\label{sec:doubleslit}
For a QBist, the double-slit experiment is about the peculiarities
that happen when an agent tries to relate their expectations for one
hypothetical scenario to their expectations for another.  Per
tradition, we can call this agent Alice.  She might compute her
probability for a detector click given that she will place the
detector at position $x$ and open slit $\#1$ --- call it
$P_1(x)$. Likewise, she can compute the corresponding quantities for
the configuration with only slit $\#2$ open, $P_2(x)$; and for when
both slits are open, $P_{12}(x)$.  All of these quantities are, by
themselves, rather ordinary probabilities: None of them end up being
negative, let alone complex.  Nor is it surprising that $P_1(x)$ might
be discrepant from $P_2(x)$ or from $P_{12}(x)$. Different conditions,
different probabilities!  The puzzle is that
\begin{equation}
  P_{12}(x) \neq P_1(x) + P_2(x).
\end{equation}
The strangeness lies not in the curve for any particular scenario, but
in how the scenarios fit together.

Rob Spekkens likes to point out that the \emph{mere fact} of
interference is not a very deep probe of quantum theory, because
interference can arise in models based on local hidden variables
\cite{Spekkens:2007, spekkens-pirsa2016}.  You just have to be careful
and consistent when constructing your model.  In his toy theory, where
states are probability distributions over discrete local hidden
variables, we can build a test for double-slit-type oddities (a toy
Mach--Zehnder interferometer), and indeed, interference occurs.

In order to test quantum theory more stringently, we have to find
probes of nonclassical expectation-meshing that resist easy
emulation. This, from the QBist perspective, is what Bell inequality
violations are all about: Given any particular choice of detector
settings, the outcome probabilities are just probabilities.  The power
and the mystery of quantum theory reside in the relation between
probabilities for different choices of detector settings.

Our research on SICs is also in this vein (see \S\ref{sec:technical}).
Using a SIC as a reference measurement is like considering a
generalized interference experiment, where the outcomes for the
``which-way'' measurement correspond to \emph{nonorthogonal} quantum
states. This generalization takes us out of the realm of easy
classical emulation, letting us investigate the quantum formalism more
deeply.

\section{Doesn't the PBR theorem prove QB\lowercase{ism} wrong?}
\label{sec:pbr}
The Pusey--Barrett--Rudolph (PBR) no-go theorem demonstrates, as the
authors put it, ``that any model in which a quantum state represents
mere information about an underlying physical state of the system, and
in which systems that are prepared independently have independent
physical states, must make predictions which contradict those of
quantum theory''~\cite{Pusey:2012}. In the years since its appearance,
many have claimed that the PBR theorem proves quantum states are ontic
--- that it rules out all epistemic and doxastic interpretations. One
often hears that QBism, having itself a doxastic conception of quantum
states, should therefore be ruled out by the lack of any experimental
violations of quantum theory.

But one should not believe these rumors. The PBR theorem does no
damage to QBism. PBR say so themselves at the end of their paper.
This is because what they demonstrate is the inconsistency of the idea
of holding epistemic quantum states at the same time as holding that
they are epistemic \emph{about} ontic states.  In QBism, quantum
states represent one's beliefs, not about some ontic variable, but
about one's future \emph{personal} experiences which come in
consequence of taking an action on the external world.  I.e., they are
epistemic (or better, doxastic) \emph{about} personal experiences.
Technically, this means there are no compelling reasons in QBism to
adopt the very starting point of PBR --- namely, trying to use an
integral over ontic states $\lambda$ to get probabilities. The PBR
theorem is a no-go result for a direction in which we never wanted to
go.

The foundational assumption of the PBR theorem is a rule for computing
some quantities $p(k|\Psi(x_1,\ldots,x_n))$, probabilities for a
measurement outcome $k$ given preparation of a product state
$\Psi(x_1,\ldots,x_n)$. This rule is a statement about conditional
probabilities:
\begin{equation}
  p(k|\Psi(x_1,\ldots,x_n))
  = \int_\Lambda\cdots\int_\Lambda
  p(k|\lambda_1,\ldots,\lambda_n)\mu_{x_1}(\lambda_1)
  \cdots\mu_{x_n}(\lambda_n)d\lambda_1\cdots d\lambda_n\,.
  \label{eq:HV}
\end{equation}
Here, $\lambda_i$ in a measure space $\Lambda$ is a possible physical
state that a system can be in, $\mu_{x_i}(\lambda_i)$ is a probability
distribution over $\Lambda$ for the $i$th system, and
$p(k|\lambda_1,\ldots,\lambda_n)$ is a probability for obtaining
outcome $k$ given a set of physical states for each system. In other
words, the whole approach of PBR is trying to identify the Born Rule
with an application of the Law of Total Probability (LTP). It can't be
done, and they have rediscovered that in their own way.\footnote{We
  note that one philosopher of physics has declared, speaking of an
  assumption equivalent to Eq.~(\ref{eq:HV}), ``If you don't believe
  that, you don't believe in physics at all.'' As best as we can tell,
  there is no reason to accept such a claim, other than an
  underdeveloped imagination.}

The LTP is familiar and what one would use \textit{if there were}
underlying hidden variables. One avenue of QBist technical research
currently ongoing is to explore an \emph{alternative} to the LTP which
expresses the fact that such hidden variables do not exist. The
crucial idea is a \emph{reference measurement,} a procedure with the
property that a probability distribution over its outcomes can be used
to compute the probabilities for all the outcomes of any other
measurement. Let $P(H_i)$ be Alice's probability for obtaining outcome
$H_i$ in an optimal reference measurement (many criteria for
optimality turn out to be equivalent for this problem). Classical
intuition suggests that the best possible reference measurement would
just be to read off the ontic state, and so by the LTP,
\begin{equation}
P(D_j) = \sum_i P(H_i) P(D_j|H_i)\,,
\end{equation}
for any other measurement $\{D_j\}$. But in the quantum world, this
does not apply, and the closest we can get to it, by cannily choosing
our reference measurement, is
\begin{equation}
    Q(D_j)=\sum_{i=1}^{d^2}\left[(d+1)P(H_i)-\frac{1}{d}\right]P(D_j|H_i)\,.
\end{equation}
$Q(D_j)$ now represents an agent's probability for obtaining the
experience $D_j$ from a measurement she represents with the POVM
$\{D_j\}$, $P(H_i)$ is her probability for obtaining outcome
$H_i$ in a hypothetical reference measurement, and $P(D_j|H_i)$ is her
probability, asserted now, for obtaining the experience $D_j$
supposing she had previously made the reference measurement and
obtained experience $H_i$. Note that the only difference from the LTP
is a constant shift and rescaling of $P(H_i)$ for each $i$. In fact,
this is the \emph{closest} \cite{DeBrota18} the two expressions can
come, suggesting that this expression may provide insight into what it
is about the universe that makes it ``quantum''.

So, in all, QBists say this about the PBR theorem (and similarly about
Bell's theorem): Rather than denigrate the QBist conception of quantum
theory, they actually help compel it.  There are so many arguments of
analogy for epistemic quantum states (Rob Spekkens' toy model nails
about 25 of them~\cite{Spekkens:2007}), but what the PBR and Bell
theorems compel and the toy theories can't is that, if quantum states
are epistemic, they cannot be epistemic about some ontic variables.
The most the PBR theorem can do is rule out a middle ground that we
are not sure anyone actually occupied in the first place.

\section{Is QBism about the Bayes rule?}
\label{sec:bayesrule}
Adopting a personalist Bayesian interpretation of probability does
\emph{not} mean treating all changes of belief as applications of the
Bayes rule.  This is shocking to some people!  And distancing
ourselves from the dogmatists who claim to follow that creed is one
reason why we prefer \emph{QBism} over ``Quantum Bayesianism''.

In the tradition of Ramsey, Savage and de Finetti, there are
consistency conditions that an agent's probability assignments should
meet at any given time, \emph{and then} there are guidelines for
\emph{updating} probability assignments in response to new
experiences.  Going from the former to the latter requires making
extra assumptions --- the two are not as strongly coupled as many
people think. The Bayes rule is not a condition on how an agent
\emph{must} change her probabilities, but rather a condition for how
she should \emph{expect} that she will modify her beliefs in the light
of possible new experiences. For this observation, we credit Hacking,
Jeffrey and van Fraassen.

Fuchs and Schack go into more detail on this point in an
article~\cite{qbist-decoherence}, and we wrote a pedagogical treatment
in a book~\cite[\S 5.1]{stacey-thesis}. 

There is a common misconception afoot that being ``Bayesian''
fundamentally means using the Bayes rule to update probabilities. For
example, the Wikipedia page that lists things named after Thomas Bayes
says that ``Bayesian'' refers to ``concepts and approaches that are
ultimately based on Bayes' theorem''. This may be historically
correct, but it is not logically correct. In the personalist Bayesian
school, we first start with the idea of quantifying beliefs and
expectations as gambling commitments. Then, we impose a consistency
condition, from which the familiar rules of probability theory
follow. The idea of \emph{updating probabilities over time} in accord with
the Bayes rule arrives rather late in this development. One must first
establish the standards for probabilities being consistent with each
other at a particular time, before invoking further considerations to
establish a scheme for \emph{changing} probabilities in response to
new experiences. Bayes' theorem \emph{is} a theorem, not an
axiom. 

The ``collapse of the wavefunction'' is analogous to, and an algebraic
variant of, Bayesian conditionalization~\cite{Fuchs:2002}. Having
recognized this, we can appreciate that it clears up a mystery (or,
perhaps better put, allows us to identify a pseudo-mystery for what it
is). But the recognition of the ``quantum Bayes rule'' was an early
step on the path to QBism, and its relevance in more recent years has
if anything been rather peripheral.

\section{What technical questions have been motivated by QB\lowercase{ism}?}
\label{sec:technical}
The development of Quantum Bayesianism, and its progressive evolution
into QBism, is a story of feedback loops between technical and
philosophical questions.

The quantum de Finetti theorem was sought and proved in order to show
there could be a meaning to the phrase ``unknown quantum state'' even
from a subjectivist perspective~\cite{q-de-finetti1}.  The Quantum
Bayesians thought that without such a theorem, a subjectivist reading
of probability in quantum theory wouldn't be possible after all.  This
theorem then outgrew its foundational origins, becoming a powerful
tool for the practical problem of analyzing the security of quantum
key distribution. A quantum de Finetti theorem for ``unknown
processes'' followed from the same motivation as that for ``unknown
states''~\cite{q-de-finetti2}.

Asher Peres pointed out that quantum states are more analogous to
probability distributions over phase space --- that is, to Liouville
density functions --- than to points in phase space. In 1995, Fuchs
followed this lead and searched for examples within Liouville
mechanics that echoed quantum theory, including the aspects of quantum
theory that had been declared uniquely nonclassical. He found that the
quantum no-cloning theorem was just one such feature: A no-cloning
theorem holds in Liouville mechanics, exactly as in the quantum case.
Trying to further refine the enquiry led to the quantum
no-broadcasting theorem~\cite{Barnum:1996}.

In 2002, Caves, Fuchs and Schack took on the question of whether or
not quantum theory implied any kinds of compatibility conditions for
disparate agents' quantum state assignments~\cite{qs-compat}.  This is
a natural question to ask, if quantum states are to be interpreted
doxastically.  The work resulted in solid theorems --- and, in a twist
whose irony has gone underappreciated, Pusey, Barrett, and
Rudolph~\cite{Pusey:2012} used one of these notions to prove the PBR
theorem. (For QBism's response to the PBR theorem, see
\S\ref{sec:pbr}.)

Another example came from trying to understand what it could mean for
quantum states to be ``disturbed by measurement'' if they are not
ontic.  Answering this led to \cite{Fuchs:1996} and \cite{Fuchs:1997},
which Fuchs later turned to the purpose of defining a threshold for
successful quantum teleportation in Jeff Kimble's
lab~\cite{Furusawa:1998}. Discussion of this point can be found in
Fuchs and Jacobs~\cite{Fuchs:2001}.

More recently, at the creative interface between conceptual and
technical matters, Fuchs and Schack have made the case that the right
way to think about decoherence is with van Fraassen's \emph{reflection
  principle}~\cite{qbist-decoherence}. We suspect that there are new
theorems to be proved in this area, in addition to the conceptual
implications (such as putting a sharper point on an old argument of
Asher Peres about when black-hole evaporation should \emph{not} be
modeled with a unitary evolution~\cite{Peres:2004}).

The most active technical topic in contemporary QBism research is the
project of reconstructing quantum theory from physical
principles. Central to this is our ongoing research into symmetric
informationally complete quantum measurements (SICs). A SIC for a $d$
dimensional Hilbert space is a set of $d^2$ pure quantum states with
equal pairwise overlaps:
\begin{equation}
    \vert\braket{\psi_i}{\psi_j}\vert^2=\frac{d\delta_{ij}+1}{d+1}\ .
\end{equation}
A uniform rescaling of these states defines a POVM which is uniquely
suited to be a ``standard quantum measurement''.

Not everyone who works on SICs is devoted to QBism.  Indeed, we gather
from conversations in hotel bars that one of the prime movers in
SIC-hunting doesn't particularly care about quantum mechanics; their
appeal as geometrical objects is enough. (Historically speaking, one
of the most closely studied SIC constructions originally flowed from
the pen of Coxeter, who just really liked
polytopes~\cite{Fuchs:2017}. But in a surprise twist, this SIC arises
in the study of quantum-state compatibility that Caves, Fuchs and
Schack initiated~\cite{Stacey:2016c}!)  Another SIC researcher is not
a QBist, but came to the problem through Fuchs's advocacy and over the
years has displayed many sympathies.

Going in the other direction, being a QBist doesn't mean you have to
live and breathe SICs.  Fuchs and BCS put it the following
way~\cite{Fuchs:2016}:
\begin{quotation}
  \noindent If all that you desire is a story that you can tell about
  the current quantum formalism, then all this business about SICs and
  probabilistic representations might be of little moment.  Of our
  fellow QBists, we know of one who likely doesn't care one way or the
  other about whether SICs exist.  Another would like to see a general
  proof come to pass, but is willing to believe that QBism can just as
  well be developed without them --- i.e., they are not part of the
  essential philosophical ideas --- and is always quick to make this
  point.  On the other hand, we two are inclined to believe that QBism
  will become stagnant in the way of \emph{all other} quantum
  foundations programs without a deliberate effort to rebuild the
  formalism.
\end{quotation}

We find that SICs cut to the heart of quantum theory in a way that
other ideas for rebuilding the formalism do not.  This is a point we
discuss elsewhere in this collection (\S\ref{sec:doubleslit}), and in
earlier papers~\cite{Fuchs:2016, qplex, Stacey:2018}.  The
representation of quantum theory that SICs furnish has natural
connections with the study of Wigner-function negativity, which is
important for quantum computation~\cite{DeBrota:2017}. In addition,
the discovery of a connection between SICs and algebraic number theory
reshapes the boundary between physics and pure mathematics in a
remarkable way~\cite{Appleby:2017, Kopp:2018}.

\section{Isn't quantum probability just classical probability but
  noncommutative?}
\label{sec:noncommute}
There's a \booktitle{Far Side} cartoon that shows a man waking up in
bed and staring at a giant note he wrote for himself on the wall:
``First pants, \emph{then} shoes!''  The lesson is that order of
operations matters in daily life, long before it matters in quantum
physics.  So, we have to be careful what we mean by ``noncommuting'',
if we want it to have any meaningful content.  And when we do get
appropriately mathematical about it, we find that it is \emph{not} the
signature of the quantum.  The Spekkens toy model, which has a simple
statement in terms of underlying local hidden variables, has
observables that do not commute~\cite{Spekkens:2007}.

There is a common sentiment about that quantum mechanics is ``a
noncommutative generalization of probability theory'': Instead of
using vectors that sum to 1, one has matrices whose trace is 1, and so
forth. This is a fine approach for many applications, but in physics,
there is never a guarantee that a method which works for one set of
problems will do equally well with another. Taking one representation
of the theory as defining its essence can cloud your physical
insight. In this case, the ``we must generalize probability to make it
noncommutative'' impulse obscures the fact that \emph{given a specific
  experimental scenario,} the probabilities of quantum physics are
just probabilities --- numbers that play together in accord with
Kolmogorov's rules. As we noted in \S\ref{sec:doubleslit}, it is the
meshing of expectations for one scenario with those of another which
reveals the fundamental enigma of quantum theory. Noncommutativity is
a secondary property, and as the Spekkens toy model teaches us, not a
quintessentially quantum one at that.

BCS, who came to QBism from statistical physics, likes to point out
that the Doi--Peliti formalism for nonequilibrium stochastic dynamics
has noncommuting operators, and also complex numbers,
Feynman diagrams, renormalization, Glauber states, the Heisenberg
equation of motion, and even the Schwinger representation of
$\mathfrak{su}(N)$.  Yet it is all a fully classical
theory~\cite{stacey-thesis, Baez:2012}. It borrows calculational
devices from quantum mechanics, but the stochasticity it considers is,
at root, ignorance about pedestrian hidden variables.

\section{Doesn't decoherence solve quantum foundations?}
\label{sec:decoherence}
The theory of decoherence is a set of calculations which enable one to
write a density matrix that is nearly diagonal in some basis of interest.
This does not tell you what a density matrix \emph{means.}

Max Schlosshauer, who wrote the canonical textbook on decoherence,
recently summarized the situation as follows~\cite{Schlosshauer:2014}:
\begin{quotation}
  \noindent Decoherence, at its heart, is a technical result
  concerning the dynamics and measurement statistics of open quantum
  systems. From this view, decoherence merely addresses a
  \emph{consistency problem,} by explaining how and when the quantum
  probability distributions approach the classically expected
  distributions. Since decoherence follows directly from an
  application of the quantum formalism to interacting quantum systems,
  it is not tied to any particular interpretation of quantum
  mechanics, nor does it supply such an interpretation, nor does it
  amount to a theory that could make predictions beyond those of
  standard quantum mechanics.

  The predictively relevant part of decoherence theory relies on
  reduced density matrices, whose formalism and interpretation presume
  the collapse posultate and Born's rule. If we understand the
  ``quantum measurement problem'' as the question of how to reconcile
  the linear, deterministic evolution described by the Schr\"odinger
  equation with the occurrence of random measurement outcomes, then
  decoherence has not solved this problem.
\end{quotation}

For a deeper dive into the QBist take on decoherence,
see~\cite{qbist-decoherence}.

\section{Is QB\lowercase{ism} like Rovelli's ``Relational Quantum Mechanics''?}
\label{sec:rqm}
Several people have made the comparison between QBism and Rovelli's
``Relational Quantum Mechanics''~\cite{Rovelli:1996}, and it is not
unjust.  Some slogans of RQM can be carried over to QBism with only a
little modification, and the \emph{motivation} for the research
program that Rovelli suggested in his original paper has certain
affinities with our own.  However, there are important differences
between QBism and RQM, and moreover, we find the statements of RQM
imprecise on key points.

Both QBism and Rovellian RQM reject the notion of a single quantum
state for the entire universe.  In QBism, measurement outcomes are
personal experiences for the agent who elicits them, while in RQM,
physical properties exist ``relationally'' between systems. As the
\booktitle{Stanford Encyclopedia of Philosophy} says, in RQM,
``Quantum events only happen in interactions between systems, and the
fact that a quantum event has happened is only true with respect to
the systems involved in the interaction''~\cite{Stanford}.  This motto
is not unlike what we have written about QBism. For example,
\begin{quotation}
  \noindent Certainly QBism has creation going on all the time and
  everywhere; quantum measurement is just about an agent hitching a
  ride and partaking in that ubiquitous process.
\end{quotation}

But we can already start to see a divergence.  Rovellian RQM downplays
the idea of agency: In RQM, a grain of sand can be an ``observer'' of
another quantum system.  Given any two systems $S_1$ and $S_2$, there
is a quantum state of~$S_2$ relative to~$S_1$, just as in Newtonian
physics, $S_2$ always has a velocity relative to~$S_1$.

Likewise, QBism and RQM differ on how to interpret probability.  While
we find the foundational papers of RQM somewhat vague on this point,
our overall impression is that RQM leans more to a Jaynesian kind of
Bayesianism, more objective and less personalist than the Ramseyian/de
Finettian school to which QBism adheres.  This is tied to a point
emphasized in the technical side of QBism (\S
\ref{sec:technical}). Mathematically speaking, a quantum state
\emph{is} a probability distribution.  Pick any informationally
complete POVM, and you can replace density operators with probability
distributions over the outcomes of that POVM (even when the density
operators are rank-1 projectors, i.e., pure states).  As best we can
tell from reading Rovelli \emph{et al.,} whenever an ``observer''
$S_1$ coexists with another system $S_2$, there exists a unique,
physically correct quantum state for $S_2$ relative to the observer
$S_1$.  Therefore, there exists a unique, physically mandated set of
probabilities concerning $S_2$, which happen to be relative to~$S_1$.
We find this philosophy of probability ultimately
untenable~\cite{Appleby:2005a, Appleby:2005b}.

We must also admit, we're not great fans of the word
\emph{relational.}  This adjective naturally carries the connotation
of ``just like in relativity theory''.  But in relativity, we can
readily transform between reference frames.  A statement like ``the
clocks $C_1$ and $C_2$ are synchronized'' is relational: Its truth or
falsity depends on whether it is evaluated by Alice or by Bob.  Yet if
Alice knows Bob's trajectory relative to herself, she can take what
she sees and Lorentz-transform her figures to compute what Bob must
see.

In quantum theory, there is no analogue of this.  (Emphasizing this
point of \emph{dis}-analogy is another way QBism distinguishes itself
from Bohr~\cite{Stacey:2018b}.) RQM tries to invent one, but the
attempt flounders.  We can see exactly how this happens if we examine
Smerlak and Rovelli's paper ``Relational
EPR''~\cite{SmerlakRovelli2007}.  The authors take a certain notion of
consistency among multiple observers over from Rovelli's original
paper:
\begin{quotation}
\noindent It is one of the most remarkable features of
  quantum mechanics that indeed it automatically guarantees precisely
  the kind of consistency that we see in nature [Rovelli 1996]. Let us
  illustrate this assuming that both $A$ and $B$ measure the spin in
  the same direction, say $z$, that is $n = n' = z$.
\end{quotation}
But on the very next page, they describe the following scenario:
\begin{quotation}
\noindent $A$ observes the spin in a given direction to be
  $\uparrow$ and $B$ observes the spin in the same direction to be
  also $\uparrow$.
\end{quotation}
And they say that \emph{this} is an ill-posed statement,
because
\begin{quotation}
\noindent it does not happen either with respect to $A$ or with
respect to $B$. The two sequences of events (the one with respect to
$A$ and the one with respect to $B$) are distinct accounts of the same
reality that cannot and should not be juxtaposed.
\end{quotation}
But if the second statement is an invalid proposition, \emph{then the
  first must be as well.}  The description ``both $A$ and $B$ measure
the spin in the same direction'' cannot apply ``either with respect to
$A$ or with respect to $B$''; it presumes a view from nowhere.  (One
could try to evade this by interpreting the story of what both $A$ and
$B$ measure as told relative to a third party, the superobserver $C$.
This might look like it could ameliorate the problem, at least if the
difficulties we saw above could be resolved.  But presuming that a
superobserver is \emph{always} available, and that the expectations of
the superobserver override those of any other participant, just
de-relationalizes the theory all over again.  And why \emph{should}
physics guarantee on a fundamental level that a superobserver is
always available?  When children or politicians quarrel, life does not
always provide a responsible adult who can restore the peace.)  In
short, the description of the gedankenexperiment that Smerlak and
Rovelli use to put forth their notion of ``consistency'' is exactly
the kind of language which they elsewhere insist is meaningless.

One philosophy paper that compared QBism and RQM~\cite{Ruyant2018}
must be mentioned in particular.\footnote{We have the sense that, like
  Bohmian mechanics, RQM has been of interest to philosophers more
  than it has been to physicists. The question of what biases the
  philosophy community perpetuates by always turning to its familiar
  authorities for opinions is an interesting one.}  We reproduce the
relevant passage with its absence of citations preserved intact:
\begin{quotation}
  \noindent QBism is the view that quantum mechanics is not a theory
  about the world, but about our degrees of credence concerning
  predictions. The theory provides universal, objective rules for
  updating these degrees from the information one gets on the world
  through events. All this is shared by RQM. One difference is that
  QBism is human-centered, while RQM is not: any physical object
  qualifies as a potential observer. But what remains of it if all
  talk of external observers boils down to talk of events relative to
  us? If anything, RQM is more radically instrumentalist than QBism:
  after all, the latter assumes that events are objective and publicly
  accessible\ldots
\end{quotation}
Most of this is at least a little wrong, so we will go through it in
detail.
\begin{quotation}
  \noindent QBism is the view that quantum mechanics is not a theory
  about the world, but about our degrees of credence concerning
  predictions.
\end{quotation}
In QBism, quantum mechanics is not a theory \emph{directly} about the
world, but rather, a theory that any of us can use to manage our
``degrees of credence'' \emph{in light of the fact that} the world has
a specific character.
\begin{quotation}
  \noindent The theory provides universal, objective rules for
  updating these degrees from the information one gets on the world
  through events.
\end{quotation}
Yes, the rules that quantum theory provides are ``universal'' (anyone
can pick up the hero's handbook~\cite{Fuchs:2016}) and ``objective''
(or as objective as anyone could want of a physical theory). The
emphasis on ``updating'' echoes a misconception we have seen
elsewhere, that Bayesian probability is fundamentally about the Bayes
update rule (see \S\ref{sec:bayesrule}). And in the QBist
understanding of personalist probability, the rules allow more loose
play in updating expectations than this formulation grants.
\begin{quotation}
  \noindent All this is shared by RQM.
\end{quotation}
To us, it seems a better fit for RQM than for QBism.  As we wrote
above, a preference for objective probability runs through RQM,
holding it back.
\begin{quotation}
  \noindent One difference is that QBism is human-centered, while RQM
  is not: any physical object qualifies as a potential observer.
\end{quotation}
\emph{Human}-centered, no, but \emph{agent}-centered, yes. An agent
does not have to be human (see sections \S \ref{sec:agent} and \S
\ref{sec:future}).
\begin{quotation}
  \noindent If anything, RQM is more radically instrumentalist than QBism:
  after all, the latter assumes that events are objective and publicly
  accessible\ldots
\end{quotation}
No, it doesn't.  Fuchs put it this way in 2010~\cite{Voldemort}:
\begin{quotation}
  \noindent \emph{Whose information?} ``Mine!'' \emph{Information
    about what?} ``The consequences (for \emph{me}) of \emph{my}
  actions upon the physical system!'' It's all ``I-I-me-me mine,'' as
  the Beatles sang.
\end{quotation}
That article goes on to draw an explicit contrast between QBism and
Pauli's claim that measurement outcomes ``are objectively available
for anyone's inspection''.

The introductory paper by Fuchs, Mermin and Schack~\cite{Fuchs:2013}
expresses the point as follows:
\begin{quotation}
  \noindent The personal internal awareness of agents other than Alice
  of their own private experience is, by its very nature, inaccessible
  to Alice, and therefore not something she can apply quantum
  mechanics to. But verbal or written reports to Alice by other agents
  that attempt to represent their private experiences are indeed part
  of Alice's external world, and therefore suitable for her
  applications of quantum mechanics. Having always stressed the
  crucial importance of stating the results of experiments in ordinary
  language, Bohr would probably have been comfortable with Alice's
  indirect access to Bob's experience, through language.

  But Bohr would not have approved of Alice superposing reports from
  Bob about his own experience, as QBism requires her to do if she
  wants to subject those reports to analysis before they enter her own
  experience. We believe that Bohr would have viewed Bob's reports ---
  formulations in ordinary language --- as beyond the scope of quantum
  mechanics. But because Alice can treat Bob as an external physical
  system, according to QBism she can assign him a quantum state that
  encodes her probabilities for the possible answers to any question
  she puts to him. When Alice elicits an answer from Bob, she treats
  this as she treats any other quantum measurement. Bob's answer is
  created for Alice only when it enters her experience. A QBist does
  not treat Alice's interaction with Bob any differently from, say,
  her interaction with a Stern--Gerlach apparatus, or with an atom
  entering that apparatus.
\end{quotation}
Or, later and more compactly:
\begin{quotation}
  \noindent What the usual story [of Wigner's Friend] overlooks is
  that the coming into existence of a particular measurement outcome
  is valid only for the agent experiencing that outcome.
\end{quotation}

\section{Why do QB\lowercase{ists} prefer de Finetti over Cox?}
\label{sec:cox}
The Cox approach is too psychologically loaded in the direction of
hidden variables and inferences about them. This sentiment dates back
to the 1990s, when Fuchs and colleagues were hashing out the basics of
being Bayesian in a quantum world.  During 1993 and 1994, Fuchs and
Schack became disenchanted with Cox's development of probability
theory and attracted instead more to the development of de Finetti and
Savage and others.  The essence of the latter school is the Dutch-book
notion and/or the simultaneous development of probabilities with
utilities (i.e., decision theory).  Looking back on it, the attraction
to the one over the other cuts to a rather fundamental point:

QBism regards physics, and science in general, in Darwinian terms.
The mathematics we develop is practical because, at root, it helps
agents to survive. From this point of view, the idea of a probability
as a gambling commitment, a belief made quantitative and ready to be
acted upon, is an attractive notion.  On the other hand, the idea of
probability being used for a ``theory of inference'' in the usual
sense --- i.e., a measure of plausibility for something that is ``out
there'' but unknown --- is a bit off-putting.

(This also seems to be a fundamental distinction between our program
and that of Rob Spekkens.  The general tenor of the Spekkensian
program has been to interpret quantum states as states of information
about some type of hidden variable as yet unspecified, perhaps degrees
of freedom that are ``relational'' in some way. The Coxian attitude is
a natural fit for this view, but it is not so for QBism.)

All the way back in July 1996, Fuchs wrote the following, in a note to
Sam Braunstein:
\begin{quotation}
\noindent While in Torino, you really got me interested in the old
          [Cox derivation] question again.  I noticed in this version
          of the book that Jaynes makes some points about how there
          are still quite a few questions about how to set priors when
          you don't even know how many outcomes there are to a given
          experiment, i.e., you don't even know the cardinality of
          your sample space.  That, it seems to me, has something of
          the flavor of quantum mechanics \ldots\ where you have an
          extra freedom not even imagined in classical probability.
          The states of knowledge are now quantum states instead of
          probability distributions; and one reason for this is that
          the sample space is not fixed --- any POVM corresponds to a
          valid question of the system. The number of outcomes of the
          experiment can be as small as two or, instead, as large as
          you want.

However I don't think there's anything interesting to be gained from
\emph{simply} trying to redo the Coxian ``plausibility'' argument but
with complex numbers. It seems to me that it'll more necessarily be
something along the lines of: ``When you ask me, ``Where do all the
quantum mechanical outcomes come from?''  I must reply, ``There is no
where there.''  [\ldots]  That is
to say, my favorite ``happy'' thought is that when we know how to
properly take into account the piece of prior information that ``there
is no where there'' concerning the origin of quantum mechanical
measurement outcomes, then we will be left with ``plausibility
spaces'' that are so restricted as to be isomorphic to Hilbert spaces.
But that's just thinking my fantasies out loud.
\end{quotation}

More recently, we have made steps in this direction, as documented in
our earlier papers~\cite{Fuchs:2016, qplex, Fuchs:2013b} and outlined
in \S\ref{sec:technical}.

\section{Why so much emphasis on finite-dimensional Hilbert spaces?}
\label{sec:finite}
Quantum theory can be formulated for finite- and infinite-dimensional
systems. By any standard, genuinely nonclassical effects are present
in finite-dimensional systems, suggesting that these may be all
that is strictly necessary for capturing the conceptual core of the
theory. Indeed, it might even be distracting to let infinite
dimensions complicate foundational considerations. In some ways the
infinite-dimensional situation is the limit of large dimensions, but
in other ways it isn't.

Infinite dimensions are subtle and complicated, but it seems they are
not so for ``quantum'' reasons.

The goal of our research is to bring clarity to the quantum mysteries.
When one looks up what the ``quantum mysteries'' are, one finds that
either they are expressed in finite-dimensional terms from the
get-go~\cite{Mermin:1981}, or, if the presentation includes continuous
degrees of freedom, all the interesting stuff happens in the
finite-dimensional part.  For example, Asher Peres' book explains a
Bell--EPR scenario using both position and spin degrees of freedom,
but the essence of the problem lies in the spins, while the position
coordinates just provide conceptual scaffolding.  To ``go for the
jugular'' of the quantum enigmas, we have chosen to focus on finite
dimensions --- and the results have been so pretty that we can't help
but wonder if they offer a guide for where physics should go
\emph{next,} as it pushes beyond the continuum theories we all know so
well.

The authors of this FAQ spend our weekdays reformulating
finite-dimensional quantum theory (see \S\ref{sec:technical}).
However, we would have nothing personal against anyone who tried to
find a new representation for, say, algebraic quantum field theory.
We do offer a cautionary note: Even the most successful and most
``fundamental'' physical theories are provisional, their applicability
contingent on physicists' limited abilities as agents to intervene
into the affairs of other natural systems.  Indeed, the way we extract
any empirical utility from a QFT is, in practice, to remind ourselves
that it \emph{cannot} be valid to arbitrarily high energies, and then
managing the consequences of that limitation (a process technically known
as regularization and renormalization).  When one cannot trust
\emph{any} physical theory to provide ultimate, metaphysical bedrock;
when \emph{all} the theories one might wish to reformulate and
reconstruct are inextricably provisional --- then, unavoidably,
picking the theory to focus upon becomes a judgment call.

It is intriguing that the possibility that physically accessible
Hilbert-space dimension is always finite --- possibly quite large, but
still finite --- is a recurring theme in quantum-gravity research. For
various flavors of this idea, see, e.g., \cite{tHooft:1993,
  Witten:2001, Parikh:2005, Banks:2018}.  Fuchs and BCS gave a QBist
spin on this speculation in 2016~\cite{Fuchs:2016}, following a lead
that Fuchs set out in 2010~\cite{Fuchs:2010} and
2004~\cite{Fuchs:2004}.\footnote{And, in correspondence with Bill
  Unruh and others, even before that~\cite[pp.\ 659--52]{Fuchs:2014}.}

\section{Aren't the probabilities in quantum physics objective?}
\label{sec:objprob}
The intuition that the probabilities in quantum physics are objective
properties of a system is deeply ingrained. For many, the suggestion
that it might be otherwise is so outlandish as to obviate the need for
rebuttal. Thus the starting point of QBism, adopting a strict, de
Finettian/Ramseyan interpretation of \textit{all} probabilities, turns out to
be a big pill to swallow once the full seriousness of its consequences
are realized. However, QBists do not deny the objective probability
intution. What we claim is that the advantages that subjectivity
brings (which may be found in any exposition of QBism) outweigh the
draw of untutored impulses. In fact, the appeal of this intuition may
be understood from and thereby absorbed into a purely personalist
point of view.

There is nothing about the intuition which demands the invocation of
quantum theory. For instance, we might just as well consider a coin or
a die. One often hears that the symmetries of the matter distribution
making up a ``fair'' coin or a die \emph{determine} the probability of
a flip landing ``heads'' or of rolling a ``3''. But what does it mean
for a coin to be ``fair''? It means that one assigns equal probability
to the heads and tails outcomes. How does one certify that a coin is
fair? If the answer involves checking that the coin's mass
distribution closely matches that of a thin cylinder, claiming that
the probability distribution comes from the mass distribution is
circular. We bring many expectations and a lifetime of experience to
the table when asserting a probability. Among these is experience with
the effects of gravity on differently shaped objects. The reason that
it feels our probabilities are properties of objects is just that we
feel the force of our priors so strongly that we feel they were given
to us by nature.

More generally, if we wanted the probability to be physically
determined, a little reflection reveals it couldn't be a property
\textit{only} of the coin itself. It must also depend on the flipping
process. A coin can have a very even mass distribution while it sits
forgotten on the bedside table. For that matter, it is quite possible
to engineer a machine which precisely flips a coin to land heads up
every time~\cite{Diaconis:2018}. Furthermore, couldn't a high-speed
camera and a sufficiently advanced computer program predict the result
of any particular coin toss with amazingly few errors given the first
few fractions of a second of the flip? With such a setup, what should
we say is the probability of heads after the machine announces its
prediction?

Supposing the force of these arguments is felt and the conclusion that
probability is about personal expectations is accepted, there remains
one refuge for the objective probabilists --- essentially, that
quantum theory legitimizes them. Classically, one might argue,
complete information is in principle possible, but quantum
mechanically, maximal information is incomplete. What's left over is
the objective chance. If one knew the objective chance, they would be
best served by setting their personal expectation equal to it.

First, we note that maximal information being incomplete doesn't require the \textit{nature} of probability to change. Supposing there is a correct probability in a given circumstance remains a big leap. But there is a more critical issue, namely, if there were a correct probability, there's no way to be sure you've got it. Here's how Fuchs and BCS put it in a previous paper~\cite{Fuchs:2016}.
\begin{quotation}
\noindent Previous to  Bayesianism, probability was often thought to be a physical property---something objective and having nothing to do with decision-making or agents at all.  But when thought so, it could be thought only inconsistently so.  And hell hath no fury like an inconsistency scorned.
The trouble is always the same in all its varied and complicated forms:  If probability is to be a physical property, it had better be a rather ghostly one---one that can be told of in campfire stories, but never quite prodded out of the shadows.  Here's a sample dialogue:
\begin{quote}
\begin{description}
\item[Pre-Bayesian:]  \ Ridiculous, probabilities are without doubt objective.  They can be seen in the relative frequencies they cause.
\item[Bayesian:]  So if $p=0.75$ for some event, after 1000 trials we'll see exactly 750 such events?
\item[Pre-Bayesian:]  \ You might, but most likely you won't see that exactly.  You're just likely to see something close to it.
\item[Bayesian:]  \ ``Likely''?  ``Close''?  How do you define or quantify these things without making reference to your degrees of belief for what will happen?
\item[Pre-Bayesian:]  \ Well, in any case, in the infinite limit the correct frequency will definitely occur.
\item[Bayesian:]  \ How would I know?  Are you saying that in one billion trials I could not possibly see an ``incorrect'' frequency?  In one trillion?
\item[Pre-Bayesian:]  \ OK, you can in principle see an {\it incorrect\/} frequency, but it'd be ever less {\it likely}!
\item[Bayesian:]  \ Tell me once again, what does ``likely'' mean?
\end{description}
\end{quote}
This is a cartoon of course, but it captures the essence and the futility of every such debate.  It is better to admit at the outset that probability is a degree of belief, and deal with the world on its own terms as it coughs up its objects and events.  What do we gain for our theoretical conceptions by saying that along with each actual event there is a ghostly spirit (its ``objective probability,'' its ``propensity,'' its ``objective chance'') gently nudging it to happen just as it did?  Objects and events are enough by themselves.
\end{quotation}

To see how quantum physics does not make probabilities somehow more objective, consider the following \cite{Fuchs:2013b}. Take a two-qubit system for which
an agents could make either of the two quantum state assignments $\rho_+$ and
$\rho_-$, defined by
\begin{equation}
    \rho_\pm=\frac{1}{2}\left(\ket{0}\bra{0}^{\otimes 2}+\ket{\pm}\bra{\pm}^{\otimes 2}\right)
\end{equation}
where we have used the common notation
\begin{equation}
    \ket{\pm}=\sqrt{\frac{1}{2}}(\ket{0}\pm\ket{1})\,.
\end{equation}
These state assignments are ``compatible'' in that they have
overlapping supports on the two-qubit state space. Yet suppose the
first qubit is measured in the ``computational basis''
$\{\ket{0},\ket{1}\}$ and outcome $1$ is found. The agent updates her
state accordingly, using the standard L\"uders rule, and her
postmeasurement state for the second qubit is then $\ket{+}$.
However, if she had begun with the joint state $\rho_-$, then
experiencing outcome $1$ would have led her to update her state for
the second qubit to $\ket{-}$ instead. The two possibilities for the
initial state were compatible, but the two possible final states,
updated in response to exactly the same data, are orthogonal! This is
an illustrative extreme case of a phenomenon that is much more
general: Priors do not inevitably wash out, even in the limit of
infinite data~\cite{priors}.

\section{Don't you have to define ``agent''?}
\label{sec:agent}
Fuchs and BCS wrote the following in an earlier paper:
\begin{quotation}
  \noindent Thinking of probability theory in the personalist Bayesian
  way, as an extension of formal logic, would one ever imagine that
  the notion of an agent, the user of the theory, could be derived out
  of its conceptual apparatus? Clearly not. How could you possibly get
  flesh and bones out of a calculus for making wise decisions?
  [\ldots\!] Look as one might in a probability textbook for the
  ingredients to reconstruct the reader herself, one will never find
  them. So too, the QBist says of quantum theory.
  \end{quotation}
This perspective is essentially that of L.\ J.\ Savage, who developed
rational decision theory in terms of ``consequences'', ``acts'' and
``decisions''~\cite{Savage:1954}, though where Savage says ``person'' we
say \emph{agent} instead.

An analogy may be helpful. In the Peano axioms for
arithmetic~\cite{Baez:2016}, the terms \emph{number}, \emph{zero} and
\emph{successor} are undefined primitives. They gain meaning by how
they play together. Seeking a more elementary meaning of those terms
within the same theory is not helpful. Instead of trying an analysis
--- in the literal sense, a ``breaking down'' --- one develops an
understanding by synthesis, by a bringing-together. The same can be
said of Hilbert's axiomitazation of geometry, in which \emph{point}
and \emph{line} are undefined primitives~\cite{Shulman:2016}.

The situation in personalist Bayesian probability is somewhat
similar. There is no way of carving up the terms \emph{gambler} or
\emph{expectation} into smaller conceptual atoms, at least not within
probability theory itself.  Personalist Bayesianism is a
\emph{synthetic theory of quantified expectations,} and there is
nothing troublesome about this.  QBism simply inherits this situation,
applying that synthetic understanding to quantum phenomena.

Just like \emph{point} and \emph{line}, or \emph{zero} and
\emph{number} and \emph{successor}, the terms \emph{agent} and
\emph{experience} gain meaning through their interplay.  Using them in
physics brings some baggage from their use in everyday speech, though
their meaning is altered --- refined, honed --- by deployment in the
more quantified setting.  This is nothing remarkable: Think of
\emph{force}, \emph{potential}, \emph{field} and so forth.

\section{Where and when does the agent end and the external world begin?}
\label{sec:future}
At a conference in 2016, Wayne Myrvold asked this:
\begin{quotation}
  \noindent Okay, help me understand this restriction of [the] scope of
  quantum mechanics you're proposing, because you're telling me I
  should only use quantum mechanics to calculate probabilities for
  outcomes of my future experiences, and that, compared to what most
  people think is the scope of the theory, is a really serious
  restriction of scope. So imagine that yesterday someone came to me
  and said, ``Wayne I want your advice on how to construct a nuclear
  waste storage facility.'' To do this I need to know about calculating
  probabilities of decays. So should I not care about any decays that
  might happen after I'm gone? Would it be a mistake to use quantum
  mechanics to calculate probabilities of radioactive decays hundreds
  of years after I'm dead?
\end{quotation}

The quantum formalism, understood as a normative criterion for an
agent's behavior, is rather agnostic about the character of the
agent. It says nothing about the agent's memory capacity, their rate
of energy consumption, how long they maintain conscious thought at a
stretch, or how quickly the molecules of their body are replaced by
food.  Looking for this kind of information in the quantum formalism
confuses the roles of agent and object.  If one is dully reductionist
and tries to specify the properties of an agent in more and more
physical detail, one will eventually be writing a many-body
wavefunction.  But any wavefunction is only meaningful as a mental
tool an agent carries to manage their expectations about something
else.

Likewise, the quantum formalism itself does not tell Alice how to
attach POVM elements to her experiences.  Instead, it is a handbook
that she can use to help herself be consistent, howsoever she sets
about mathematizing her life.  The formalism does not care whether she
believes that she will die tomorrow, whether she thinks she can
cryogenically freeze herself and wake up on Mars a thousand years from
now still essentially Alice, whether she regards potential genetic
descendants of herself as sharing in her good or ill fortune ---
nothing of the sort.  Instead, the formalism helps her gamble
consistently, using whatever beliefs she currently has about such
matters.

In the case of a gamble with consequences beyond an individual's
expectations for their own longevity, the ``agent'' making the bet may
be a community, rather than a single human being. Perhaps it is a
collaboration of a number of scientists which grows or shrinks as
years go by. The situation is similar to that of an individual buying
life insurance. Why would anyone ever do this? Life insurance pays out
only if the individual making the purchase dies --- it's impossible
for anyone to reap the benefits of their own life insurance
policy. The answer is quite intuitive: Because they consider their
family to be an extension of themselves. Even though they, personally,
will be gone, a conceptual part of themselves remains which can cash
the check. It is like the couple who shares a bank account and makes
purchasing decisions on the basis of ``us'' rather than either of them
alone. The concept of an agent is extremely flexible.

Quantum theory tells us that an agent can express her expectations in
terms of probabilities for a hypothetical ``Bureau of Standards''
experiment (see \S\ref{sec:pbr} and \S\ref{sec:technical}).  The BoS
experiment might be exceedingly difficult to carry out: Perhaps it
costs a hundred million dollars in optical equipment. But, even though
Alice does not physically perform it, it is mentally useful for her in
her cogitations.

What about an experiment that requires a forbiddingly large investment
of another resource --- not money, but \emph{time?}  The same binding of
expectations between different hypothetical scenarios should still
apply.  Mathematically, all the prolongation implies is an orthogonal
transformation of her probability vector.

To push it a step further: What if Alice contemplates the hypothetical
experiment of extending her own life radically?  She sees no ready
path to doing so, but she lets her imagination wander.  Could she
replace her neurons one by one with nanomachines?  Does her overall
mesh of beliefs about her own agenthood permit the idea that any
meaningful aspect of her could persist?  Even if Alice finds the whole
notion exceedingly implausible, can she treat it simply as another
experiment that would require a large resource investment to realize?

The QBist answer is ``Yes'' --- or, more carefully, that nothing in the
quantum formalism itself forbids it.

We are reminded of a lesson from a colleague.
\begin{quotation}
  \noindent This is a good example of the primary point of Dirac
  notation: it has many built in ambiguities, but it is designed so
  that any way you chose to resolve those ambiguities is correct. In
  this way elementary little theorems become consequences of the
  notation. Mathematicians tend to loathe Dirac notation, because it
  prevents them from making distinctions they consider
  important. Physicists love Dirac notation, because they are always
  forgetting that such distinctions exist and the notation liberates
  them from having to remember.

--- N.\ David Mermin, ``Lecture Notes on Quantum Computation'' (2003)
\end{quotation}

The philosophy of personal identity is brimming with ambiguities, but
living in accord with the normative principles of the quantum
formalism means that any way I choose to resolve them is correct.

\section{Does QB\lowercase{ism} lose the ``explanatory power'' of other
  interpretations?}
\label{sec:explain}
In the philosophy of science, explanations can be causal,
unificationist, deductive-nomological, statistical relevantist,
inducto-statistical, asymptotic and probably other types
besides~\cite{Weatherall2011}.  Sometimes, epochal progress is made by
declaring that an entire genre of attempted explanations is
unnecessary, misguided and counterproductive. We've been doing that
ever since some clever ancient Greek decided that they could
contemplate thunder without drawing the family tree of the
Thunderer. While Descartes pictured the planets as being dragged about
in a material whirlpool, Newton declared, ``I feign no hypotheses''
and gave us classical mechanics. The manifold complexities of living
beings did not require central planning --- only, as Darwin taught us,
heredity and luck.  Einstein postulated the constancy of the speed of
light, without worrying about how moving through the ether might
elastically deform the electron, and that is why we learn Lorentz's
equations but with Einstein's motivation.

In a sense, Newton explained \emph{less} than Kepler did, because
Kepler had a reason why there were six and only six planets: After six
planets, we run out of Platonic solids. We can rightly reject Kepler's
explanation, even in the absence of a complete story about how the
solar system happened --- and even though Newton's explanation was, by
the standards of his time, frankly un-``physical''.\footnote{Kepler's
  image of nested spheres and regular solids seems absurdly
  numerological today, though anyone who has wanted $E_8$ or the
  Monster group to appear in fundamental physics, just for the
  {\ae}sthetics of it, should feel the tug of the Platonic solids! (We
  strongly doubt that there is any ``theory of everything'' inside
  $E_8$~\cite{Distler:2010}, although the corresponding lattice does
  turn out to involve a peculiarly nice quantum
  measurement~\cite{Stacey:2017, Stacey:2016}.) Kepler's geometrical
  model was wrong, but it was specific, quantitative, directly
  inspirational and, unlike many bits of our scientific
  heritage~\cite{Bouie:2018, Switek:2019}, not breathtakingly racist,
  which maybe counts for something.}  Quantum physics leads us to go
further than Newton. Instead of merely saying ``I feign no
hypothesis'', we can declare that \emph{the character of the natural
  world is such that ``feigning a hypothesis'' --- erasing agency and
  telling a story from a God's-eye perspective --- is a bad idea.}
This is an affirmative statement about ontology, and the furthest
thing possible from asserting that the world vanishes when I close my
eyes (see \S\ref{sec:solipsism}).

To ask quantum theory for a story about what happens at the slits of a
double-slit experiment ``when nobody is looking'' is like taking
thermodynamics and saying, ``OK, but where is the phlogiston?'', or
seeing the inverse-square law of gravity and demanding to be shown the
dodecahedron that makes it go.

One motivation for the technical side of QBism (see
\S\ref{sec:technical}), particularly the project of reconstructing
quantum theory from physical principles, is to elevate the quality of
explanations of which quantum physics is capable.  The quantum
formalism can be applied to any physical system, minuscule or vast,
and so any lesson gleaned from the formalism itself must be a very
general one --- a \emph{why} that pertains, in some measure, anywhere.
We physicists tend to like explanations that cut to the fundamental
principles of a subject, particularly with a dramatic twist that makes
the argument more obvious in retrospect.  The opaque
nature of the textbook quantum formalism doesn't just make teaching
the subject difficult.  (``Master these fifty pages of differential
equations and operator theory. Just trust us. Yours not to question
why.'')  It also buries the enigmatic features of the theory, like the
violation of Bell inequalities, and limits physicists' abilities to
devise \emph{good} explanations. We aim to fix this --- but
\emph{that} is a whole project (\S\ref{sec:technical}).

When critics have challenged us on the issue of QBism's ``explanatory
power'', the type of explanation they've often had in mind is something like
what solid-state physics has to say about matter being solid.
\emph{Pauli exclusion keeps you from falling through the floor;
  checkmate, QBists!}  And in fairness, this does sound rather removed
from the scenarios that the QBist literature has mostly dwelled upon
--- an example of QBism showing its ancestry in quantum information
science.  Where are agents and interventions in the topics preferred
in solid-state society?

In physics, an explanation is not a statement made in isolation. We do
not just say, ``That rock will sit there without collapsing in on
itself.'' We naturally go a step further: ``That rock will resist
being squeezed.'' Squeezing a rock in one's hands \emph{is a quantum
  measurement} --- merely a very imprecise one, for which the
textbooks don't say much about representing by a POVM. When we invest
meaning in words like \emph{solid} and \emph{rigid} and
\emph{incompressible,} we are, at least tacitly, making claims about
how a physical system will react against interventions.  And thus,
even in solid-state mechanics, agenthood was there all along.  The
fact that we do not make single predictions in isolation is ultimately
baked into the formalism, because asserting a quantum state assignment
$\rho$ for a system implies quantitative expectations about the
outcomes of \emph{any} experiment that one can represent in the
theory. No expectation value stands alone.\footnote{When our
  measurements are sloppy, we can typically get by without the full
  apparatus of quantum theory to guide our actions. We can use dodges
  like average densities of energy levels. We can cheat and model a
  phenomenon as a classical stochastic process with mundane parameters
  like average reaction rates. The more closely we interrogate the
  world, the more we need quantum theory in order to prosper in
  it. Freedom to intervene, and precision of intervention, are
  \emph{resources.} When an agent is limited in these regards, the
  full vitality of quantum phenomena is denied them.}

\section{Aren't probabilities an insufficient representation of beliefs?}
\label{sec:insufficient}
We don't claim that personalist Bayesian probability theory is the end
of the story.  We only hold that it is adequate where it is needed: It is
a tool applicable when experiments can be defined quantitatively and
the sample spaces of their potential outcomes tabulated in advance.
BCS notes, ``This is one reason why I say \emph{expectation} instead
of \emph{belief} sometimes. It carries a bit of a connotation of
belief quantified and rigorized, rather than left raw. Plus, the
\emph{X} makes it sound cool.''

A great amount of confusion has been stirred up by the misconception
that personalist Bayesianism presumes that living human beings
actually do act as perfectly rational expectation-balancing agents.
In this regard, we share a wry observation of Diaconis and
Skyrms~\cite{Diaconis:2018}:
\begin{quotation}
  \noindent In a large and growing experimental literature in
  psychology and behavioral economics, it appears that \emph{almost
    all theories are systematically violated by some significant
    proportion of the population.} It also appears that there are
  different types in the population. Some violate one principle; some
  violate another. And there are even some expected utility
  maximizers.
\end{quotation}
In other words, the theory of personalist Bayesian probability is
normative, not descriptive.

\section{What does unitary time evolution mean in QBism?}
\label{sec:unitary}

This is a point that we addressed a bit tersely in earlier
publications~\cite{Fuchs:2016, Stacey:2016b, Fuchs:2002, Fuchs:2013b},
and which we approached from multiple directions amid a large samizdat
of miscellany~\cite{Fuchs:2014}.  In this section, we will attempt a
balance between these two levels of verbosity.

Fuchs pointed out some time ago that the arguments for the
subjectivity of quantum states also apply to unitary time
evolutions~\cite{Fuchs:2002}.  Unitaries can be toggled from a
distance; they can be teleported.  More recent work on the
probabilistic representation of quantum theory makes the point even
more directly: Quantum states are probability distributions
(\S\ref{sec:pbr}), and unitaries are \emph{conditional probabilities},
used in a way that respects the nonexistence of hidden variables.

Consider an agent Alice, who uses quantum theory to help herself
navigate the world. Accordingly, she carries a probability
distribution for an informationally complete measurement, which she
uses to summarize her expectations.  Alice can calculate other
probability distributions from it, including distributions for other
informationally complete measurements which she might carry out in the
distant future. The textbook way of writing a unitary evolution is to say
\begin{equation}
  \rho' = U(t) \rho U(t)^\dag,
\end{equation}
where the operator $U(t)$ models the passage of an amount of time
$t$. Both density operators $\rho$ and $\rho'$ express beliefs that
Alice holds \emph{now.}  The former encodes her present beliefs about
a reference measurement she might perform immediately, while the
latter encodes her beliefs about what she might experience were she to
instead perform that reference measurement at a later time. All of
these beliefs, which she expresses quantitatively as gambling
commitments, are commitments she makes at the present time.  If time 0
is Monday at noon, and time $t$ is noon on Tuesday, then $\rho$ is
Alice's gambling commitment about a measurement she might perform on
Monday, and $\rho'$ is the commitment she holds simultaneously about a
measurement to potentially be done on Tuesday.  The unitary operator
$U(t)$ is, likewise, a belief that she holds as part of the same mesh
of expectations that includes $\rho$ and $\rho'$.  It is, in this
sense, a statement \emph{synchronic} with $\rho$ and $\rho'$.  It does
\emph{not} express how Alice's beliefs must necessarily change as time
passes, though if Tuesday rolls around and Alice has not yet performed
a measurement, she can adopt her old numbers $\rho'$ as her new
expectations for a reference measurement.

For simplicity, let's assume that the system Alice is contemplating
experiments upon is one for which she knows a SIC.  She represents her
quantum state $\rho$ using the Born Rule as
\begin{equation}
  P(H_i) = \frac{1}{d} \tr(\rho \Pi_i)\,,
\end{equation}
where the projectors $\{\Pi_i\}$ satisfy
\begin{equation}
  \tr(\Pi_i \Pi_j) = \frac{d\delta_{ij} + 1}{d+1}.
\end{equation}
Then, she can write the Born Rule for any other POVM $\{D_j\}$ as a
simple modification of the Law of Total Probability:
\begin{equation}
    Q(D_j)=\sum_{i=1}^{d^2}\left[(d+1)P(H_i)-\frac{1}{d}\right]P(D_j|H_i)\,.
\end{equation}
If Alice uses something other than a SIC as her reference measurement
$\{H_i\}$, the formula will be more complicated, but the concepts are
the same. A unitary transformation of $\rho$ can be shifted onto
the elements of the reference measurement, since
\begin{equation}
  P'(H_j|t) = \tr(\rho' H_j) = \tr[U(t) \rho U^\dag(t) H_j]
  = \tr[\rho (U^\dag(t) H_j U(t))]\,.
\end{equation}
Expressing $\rho$ in terms of $P(H_i)$, this becomes
\begin{equation}
  P'(H_j|t) = \sum_{i=1}^{d^2}\left[(d+1)P(H_i)-\frac{1}{d}\right]
  R(H_j|H_i,t)\,,
\end{equation}
where
\begin{equation}
  R(H_j|H_i,t) = \frac{1}{d}\tr(U(t) \Pi_i U^\dag \Pi_j)
\end{equation}
is a matrix of conditional probabilities. The value of $R(H_j|H_i,t)$
is Alice's probability for obtaining the $j$th outcome of a reference
measurement when her state assignment is the projector $\Pi_i$,
unitarily rotated by $U(t)$. Note that classically, we would express
the relation between Alice's current probabilities for a measurement
now and her current probabilities for a measurement later as
\begin{equation}
  P'(H_j|t) = \sum_{i=1}^{d^2} P(H_i) R(H_j|H_i,t)\,.
\end{equation}
In quantum mechanics, we cannot think of time evolution as shifts in
the values taken by hidden variables, so we do not use this
expression, but rather its quantum replacement, which simplifies to
\begin{equation}
  P'(H_j|t) = (d+1) \sum_{i=1}^{d^2} P(H_i) R(H_j|H_i,t) - \frac{1}{d}\,.
\end{equation}
All of the quantities $P(H_i)$, $P'(H_j|t)$ and $R(H_j|H_i,t)$ are
beliefs that Alice holds simultaneously. They all have the same
status, in that they are personalist Bayesian probabilities, every one
of them.

Over the years, we have noticed that some people who are on board with
quantum states being subjective still balk at the prospect of
regarding quantum \emph{operations} that way. Imbuing unitaries with
subjectivity, they fear, risks the whole Standard Model going up in a
puff of arthouse smoke. This concern is understandable, but
misplaced. We find that a personalist Bayesian take on unitaries, and
the ``all QFTs are effective QFTs'' ethos of weekday field
theory~\cite{ZinnJustin:2007}, meet quite nicely if we only let
them. What follows is speculation for the future development of
physics, guided by the lessons of practical applications.

Let's suppose that Alice is a physicist who is preparing to do an
experiment, say on a spin system. Following ordinary procedure, she
writes down a unitary time-evolution operator generated by a
Hamiltonian. What does this Hamiltonian encode? Well, it expresses
what Alice is doing with her laboratory equipment: the $\vec{E}$ and
$\vec{B}$ fields established by charged capacitor plates and
current-carrying wires, for example. An old book might have called
this information ``a complete description of the apparatus in everyday
language, suitably augmented with the concepts of classical
physics''. But Alice knows that she can treat any item of her
laboratory apparatus as a quantum system in its own right. For
instance, she can use the quantum theory of solids to explain why she
can force a current through her coiled wire. So, that which she
expresses as a unitary operator, she also recognizes from a broader
perspective as a mathematical consequence, in principle, of a quantum
state assignment. The ``effective unitary'' she implements naturally
has, therefore, the same physical status as her quantum-state
ascriptions: They are all, at root, personalist Bayesian expectations.

What, then, of the most ``fundamental'' time evolutions of all? Let us
go all the way, or at least as far as modern physics can take us. What
is the status of the Standard Model Lagrangian in QBism? Apart from
the last two words, this is a question already generations old; the
project of grand unification, seen as quite respectable, has with game
persistence tried to understand the Standard Model as the low-energy
limit of a new theory, not too dissimilar to it in basic
conceptions.

We do not want to prejudge the matter and ennoble some
part of a theory too rashly. After all, human mathematicians have yet
to express a nontrivial QFT in a way that meets even their own
standards of rigor, let alone a way that would be suitable for the
``eyes of God''. Among the manifold interesting complications is the
fact that not all QFTs are written in terms of a Lagrangian, and it is
conceivable that not all of them can be~\cite{nlab:2017}.

In order to wring practical numbers out of a QFT, one admits that the
theory only applies up to some high-energy or short-distance cutoff,
and then one deals with this limitation in an emotionally mature
manner. This discipline is known as regularization and
renormalization~\cite{Baez:2016b}. A scattering amplitude is computed
as a function, not just of particle momenta and coupling strengths,
but also of the ultraviolet cutoff. Changes in some of these
parameters can be absorbed by changes in others, leaving the
scattering amplitude numerically unchanged. The theory is not just a
single choice of terms and coefficients, but the entire
renormalization-group flow.

Seen in this light, the core of a QFT begins to take on a role akin to
the Born Rule:\ a normative constraint relating expectations for
different experiments. The story of integrating over UV degrees of
freedom, beta functions, the running of coupling ``constants'' --- it
brings the message that gambles at one energy ought to be tied with
gambles at another.\footnote{A related hint comes from lattice gauge
  theory, where the gauge group is specified at a quite primitive
  level of setting up the problem~\cite[\S VII.1]{Zee:2010}, much like
  the selection of Hilbert-space dimensionality in quantum
  computation, and what follows is rather like a complexified
  MaxEnt~\cite{Baez:2013}.}

It is conceivable that when the foundational unscrambling
(\S\ref{sec:interpret-prob}) is complete, unitary operators will join
quantum states on the doxastic side of the line, while the
\emph{fundamental core} of a ``grand unified'' theory will, like the
Born Rule, reveal itself as an empirically motivated, normative
addition to probability theory.

\section{What do the recent Extended Wigner's Friend thought-experiments imply for QBism?}
\label{sec:wigner}

The past few years have seen the birth of a mini-field, where the
thought-experiment called Wigner's Friend is wrapped around a
no-hidden-variables argument~\cite{Frauchiger:2017,Brukner:2017,Pusey:2018,Bub:2018}.

We've read the papers, we've been to a workshop~\cite{workshop:2019},
and we're still not convinced that the introduction of additional
friends, robots or Wigners goes beyond the original Wigner's Friend
``paradox'' that QBism already answered on its own
terms~\cite{Fuchs:2016}. In order to deserve attention, a ``paradox''
should reveal an actual inconsistency following from the premises of
some interpretation. From our perspective, every new variant just does
other things that we know are fallacious:\ treating unitaries as
ontic, acting like systems have quantum states when there is no agent
to assign them, pretending that probabilities follow from frequencies,
making believe that decision theory is meaningful when agents are so
constrained they can make no choices, etc. The extra complexity
introduces additional opportunities for confusion, without making the
argument more forceful. We would \emph{like} this situation to change;
for example, it would be interesting to derive a new quantitative
criterion of classicality from these considerations.  But the
mini-field that studies Wigner's friends, cousins and former roommates
is not quite there yet. So far, the conditions deduced from these
thought experiments have been Bell-type inequalities given slightly
rephrased justifications, and thus they have not pointed to
fundamentally novel issues.

\section{Is QBism compatible with the Many Worlds Interpretation?}
\label{sec:mwi}
Sometimes, when ideas are presented as going off in two
opposite directions, the reason is that they really are, and there
isn't any secret centrist wisdom in trying to yoke them back together.

There is no one single Everettian faith, any more than there is truly
a unified ``Copenhagen Interpretation'' (see
\S\ref{sec:copenhagen}). Instead, the genus has many species,
frequently incompatible with one another~\cite{Kent:2015}. On rare
occasions, an apostle of one of these creeds might make a statement
that, in isolation, has a vague affinity to a QBist position. That
much is to be expected, since we are all talking about quantum
physics, and we are not trying to hang a bag of hidden variables on
the side of it (as, say, the Bohmians are wont to do). But we QBists
have no physical state vector for the entire universe, no All-Function
evolving unitarily in the eye of God.

Imagine, if you can, a physical state vector for the entire cosmos
$\ket{\Psi}$, and a factorization of the cosmic Hilbert space into
distinguished subsystems. (An Everettian creed will either presume
this or attempt to derive it, generally by way of an argument that
turns out to be circular.)  Now, pick one of those subsystems and take
the trace of $\ket{\Psi}$ over all the others.  The marginal state of
the focal subsystem is then the unique, physically mandated density
operator for that subsystem, fixed by ontology. But in QBism, there is
no such thing.

The same holds true if one tries to decompose the All-Function into
``relative states'' of observers and observed.  When the carving is
all done, the pieces are each physically mandated, ontologically fixed ---
and that's simply not the role that any quantum state plays in QBism.

A typical move for modern Everettians is to take the
quantum-mechanical formalism, chop off the Born rule and then claim to
re-derive it. Generally, the algebra can be made to cough up a set of
numerical weights, but the identification of those weights as
probabilities in any meaningful sense turns out rather
unwarranted.

Take another look at the infrastructure underlying the Everettian
story: complex Hilbert space, time evolution as unitary operator,
etc. To us, \emph{all} of those cry out for explanation. Indeed, the
Born Rule, the very part of the theory that Everettians wish to excise
--- the part to be re-derived as a technicality, delegated to the
afterthoughts --- may be the most important part of all. Properly
formulated, it might well bring the essential enigma of the quantum
into the spotlight with a clarity never before achieved~\cite{qplex}.

By contrast, we see nothing in the Everettian picture that is uniquely
compelled by quantum theory specifically. For instance, you
\emph{could} invent a Many-Worlds Interpretation of Spekkens' toy
model (as John Smolin once admitted~\cite[p.\ 1407]{Fuchs:2014}). The
result would be baroque and contrived, revealing nothing about the
model itself.

We suspect that the appeal of multiverse imagery has more to do with
psychology than with physics.  Quoting a letter Fuchs wrote in
2002~\cite[p.\ 347]{Fuchs:2014}:
\begin{quotation}
\noindent What I find egocentric about the Everett
point of view is the way it purports to be a means for us little
finite beings to get outside the universe and imagine what it is
doing as a whole.  And what is it doing as a whole?  Something
fantastic?  Something almost undreamable?!  Something inexpressible
in the words of man?!?!  Nope.  It's conforming to a scheme some guy
dreamed up in the 1950s.

This whole fantastic universe can be boiled down to something
representable within one of its most insignificant components --- the
brain of man.  Even toying with that idea, strikes me as an
egocentrism beyond belief.  The universe makes use of no principle
that cannot already be stuffed into the head of an average PhD in
physics?  The chain of logic that leads to the truth of the
four-color theorem (apparently) can't be stuffed into our heads, but
the ultimate operating principle for all that ``is'' and ``can be''
can?
\end{quotation}

Other varieties of multiversitarianism also leave us unmoved. To adapt
a line of Martin Gardner, observable universes are not even as common
as two blackberries. Proclamations about ``the multiverse'' appear to
us like failures of imagination, wrapped up in extravagances that
provide a certain unsubtle, bulk-rate imitation of it. Our cynical
view of these proclamations may be due to our preference for the
philosophy of pragmatism.\footnote{It also seems to us that arguments
  in this area tend to disconnect from actual scientific progress. For
  example, it is a genre convention to quote Weinberg's ``prediction''
  of a small, nonzero cosmological constant from anthropic
  reasoning~\cite{Ellis:2017}. Varying one parameter in isolation ---
  a parameter that we have no good reason to consider
  fundamental~\cite{Wang:2017}, at that --- while holding all others
  fixed strikes us as having dubious physical relevance. Moreover,
  Weinberg's calculation requires as input the maximal observed
  redshift of a galaxy~\cite{Nadathur:2014}. His formula coughed up a
  decent answer when this was $z = 4.4$, but it fares dramatically
  worse now that we have seen a galaxy at $z =
  11.1$~\cite{Oesch:2016}. Weinberg's argument now gives a bound on
  the vacuum energy density of about 5800 times the present cosmic
  mass density. This is three orders of magnitude larger than the
  observed value, a ratio well into the regime where Weinberg himself
  says the cosmological constant would be ``so small that even the
  anthropic principle could not explain its
  smallness''~\cite{Weinberg:1987}.}

Most likely, we are doing ourselves few favors in the pop-science
media by taking this position, but we are willing to be cast as the
stodgy ones.

As for the high-flying speculations of the ``all mathematical
structures are physically real'' variety, we find that an observation
by the philosopher William James rather encapsulates our sentiments.
The quote that follows is from a 1906 lecture.  While a modern
multiversitarian would use newer terminology, it boils down to nothing
essentially different from the ``Absolute'' and the ``mind of God''
that had taken hold of the ``rationalists'' at the time.
\begin{quotation}
\noindent The more absolutistic philosophers dwell on so high a level
of abstraction that they never even try to come down.  The absolute
mind which they offer us, the mind that makes our universe by thinking
it, might, for aught they show us to the contrary, have made any one
of a million other universes just as well as this.  You can deduce no
single actual particular from the notion of it.  It is compatible with
any state of things whatever being true here below. [\ldots\!]
Absolutism has a certain sweep and dash about it, while the usual
theism is more insipid, but both are equally remote and vacuous.
\end{quotation}

\section{What are good things to read about QB\lowercase{ism}?}
\label{sec:others}
While we're quoting William James, it's a good time to share a remark
from his \booktitle{Pragmatism} (1907), which by itself is enough to
elevate him to the first rank of intellectuals:
\begin{quotation}
  \noindent Whatever universe a professor believes in must at any rate
  be a universe that lends itself to lengthy discourse.
\end{quotation}
Accordingly, there is no shortage of primary sources about QBism. The
essay by Fuchs, Mermin and Schack in the \booktitle{American Journal
  of Physics} introduces the interpretation with an emphasis on how it
gives meaning to the standard mathematical formulation of quantum
theory~\cite{Fuchs:2013}. Mermin~\cite{Mermin-Bell, Mermin:2018} and
Fuchs~\cite{notwithstanding, Fuchs:2012} have both written pieces that
go more in depth on the historical setting of QBism.  Of these essays,
Fuchs's explains more of the technical side of current research.
Additional details of that technical work are presented
in~\cite{Fuchs:2016, Fuchs:2013b}. Fuchs also discusses the genesis of
QBism in the introduction to the samizdat
compilation~\cite{Fuchs:2014}. Pretty much every question we have
received about QBism is \emph{addressed} somewhere
in~\cite{Fuchs:2014}, though not always \emph{answered} with finality
--- QBism is, as we said above, a project.

As for secondary sources, the \booktitle{Stanford Encyclopedia of
  Philosophy} has a pretty good article on QBism and related
interpretations:
\begin{quotation}
\url{https://plato.stanford.edu/entries/quantum-bayesian/}
\end{quotation}
This was written by Richard Healey, who is not a QBist but has an
interpretational attitude that is in many ways QBism-adjacent. Being
written for an \booktitle{SEoP} audience, it is heavier on the
philosophical matters and gives less time to the technical research
that those matters have motivated.

If you want a whole book that you can carry around, Hans von Baeyer's
\booktitle{QBism: The Future of Quantum Physics} (Harvard University
Press, 2016) is an accurate portrayal, pitched to the
interested-layperson audience.

(And incidentally, on the topic of books, Persi Diaconis and Brian
Skyrms recently released \booktitle{Ten Great Ideas about Chance},
which lays out a school of thought about probability that is pretty
much aligned with the one QBism adopts. Diaconis and Skyrms confine
the quantum stuff to a single chapter, but they do recommend a David
Mermin essay on QBism as good reading~\cite{Diaconis:2018}.)

QBism has been written up both in \booktitle{New
  Scientist}~\cite{Chalmers2014} and in \booktitle{Scientific
  American}~\cite{VonBaeyer2013}, though not terribly accurately in
either case, thanks to the editorial process~\cite{Mermin14b,
  Mermin-Vienna, Mermin-Bell}.  A better treatment, albeit in German,
appeared in the \booktitle{Frankfurter Allgemeine
  Sonntagszeitung}~\cite{VonRauchhaupt2014}. \booktitle{Nature}
addressed it briefly in the context of information-oriented
reconstructions of quantum theory~\cite{Ball2013}.

In June 2015, the pop-science website \booktitle{Quanta Magazine} ran
an interview with Fuchs~\cite{Gefter2015}.  The accompanying profile
is largely accurate, except for a figure caption that implies QBism is
a hidden-variable theory:
\begin{quotation}
\noindent A quantum particle can be in a range of possible
states. When an observer makes a measurement, she instantaneously
``collapses'' the wave function into one possible state. QBism argues
that this collapse isn't mysterious. It just reflects the updated
knowledge of the observer. She didn't know where the particle was
before the measurement. Now she does.
\end{quotation}
A better caption would go more like the following:
\begin{quotation}
\noindent In the textbook way of doing quantum physics, a quantum
particle has a ``wave function'' that changes smoothly when no one is
looking, but which makes a sharp jump or ``collapse'' when the
particle is observed. QBism argues that this collapse isn't
mysterious. It just reflects the altered expectations of the
observer. Before the measurement, she didn't know what would happen to
her when she interacted with the particle. After the measurement, she
can update her expectations for her future experiences accordingly.
\end{quotation}
Originally, the subhead was also misleading; soon after the interview
appeared, \booktitle{Quanta} fixed the subhead, but not the figure
caption.  So it goes.

Later, Fuchs was interviewed for the Australian Broadcasting Company's
program, \booktitle{The Philosopher's Zone}~\cite{philzone}.

\section*{Acknowledgements}
JBD was supported in part by the Foundational Questions Institute Fund
on the Physics of the Observer (grant FQXi-RFP-1612), a donor advised
fund at the Silicon Valley Community Foundation. BCS was supported by
the John Templeton Foundation, under grant 61098, ``Geometric
Phases and Symmetric Quantum Measurements''. The opinions expressed in
this publication are those of the authors and do not necessarily
reflect the views of the John Templeton Foundation. We thank Chris
Fuchs for lengthy discussions during the writing and rewriting of this
collection.


\begin{thebibliography}{999}

\bibitem{Mermin:2013} N.\ D.\ Mermin, ``QBism as CBism: Solving the
  Problem of `the Now','' \arxiv{1312.7825} (2013).
  
\bibitem{Gefter2015} A.\ Gefter and C.\ A.\ Fuchs, ``A Private View of
  Quantum Reality,'' \booktitle{Quanta Magazine} (4 June
  2015). Available at
  \url{https://www.quantamagazine.org/20150604-quantum-bayesianism-qbism/}.

\bibitem{Fuchs:2016} C.\ A.\ Fuchs and B.\ C.\ Stacey, ``QBism:
  Quantum theory as a hero's handbook.''
  \booktitle{Proceedings of the International School of Physics
    ``Enrico Fermi,'' Course 197 -- Foundations of Quantum Physics},
    edited by E.\ M.\ Rasel, W.\ P.\ Schleich, and S.\ W\"olk
  (IOS Press, 2019). \arxiv[quant-ph]{1612.07308}.
  
\bibitem{Fuchs:2013} C.\ A.\ Fuchs, N. D. Mermin and R.\ Schack,
  ``\hrefdoi{10.1119/1.4874855}{An introduction to QBism with an
  application to the locality of quantum mechanics},''
  \booktitle{American Journal of Physics} \textbf{82} (2014), 749--54,
  \arxiv{1311.5253}.
    
\bibitem{Frauchiger:2017} D.\ Frauchiger and R.\ Renner,
  ``\hrefdoi{10.1038/s41467-018-05739-8}{Quantum theory cannot
  consistenty describe the use of itself},'' \booktitle{Nature
  Communications} \textbf{9} (2018), 3711, \arxiv{1604.07422}.

\bibitem{Brukner:2017} \v{C}.\ Brukner, ``On the quantum measurement
  problem,'' \arxiv{1507.05255}. In \booktitle{Quantum [Un]Speakables
    II} (Springer-Verlag, 2017).

\bibitem{Gardner83} Martin Gardner, ``Why I Am Not a
 Solipsist,'' in {\sl The Whys of a Philosophical Scrivener} (Quill, 1983).

\bibitem{DeBrota18} J.\ B.\ DeBrota, C.\ A.\ Fuchs and B.\ C.\ Stacey,
  ``Symmetric Informationally Complete Measurements Identify the
  Essential Difference between Classical and Quantum,''
  \arxiv{1805.08721} (2018).

\bibitem{Atmanspacher:2014} H.\ Atmanspacher, ``20th century variants
  of dual-aspect thinking,'' \booktitle{Mind and Matter} \textbf{12}
  (2014), 245--88.
 
\bibitem{Mermin14} N.\ D.\ Mermin,
  ``\hrefdoi{10.1038/507421a}{QBism puts the scientist
  back into science},'' \booktitle{Nature} {\bf 507} (2014): 421--23.

\bibitem{Stacey:2018b} B.\ C.\ Stacey, ``Misreading EPR: Variations on
  an incorrect theme,'' \arxiv{1809.01751} (2018).
    
\bibitem{Mermin:2018} N.\ D.\ Mermin,
  ``\hrefdoi{10.1088/1361-6633/aae2c6}{Making better sense of quantum
  mechanics},'' \booktitle{Reports on Progress in Physics} \textbf{82}
  (2018), 012002, \arxiv{1809.01639}.
    
\bibitem{notwithstanding} C.\ A.\ Fuchs, ``Notwithstanding Bohr, the
  Reasons for QBism,'' \booktitle{Mind and Matter} \textbf{15} (2017),
  245--300, \arxiv{1705.03483}.

\bibitem{Bohm:1951} D.\ Bohm, \booktitle{Quantum Theory}
  (Prentice-Hall, 1951).

\bibitem{Stacey:2016b} B.\ C.\ Stacey,
  ``\hrefdoi{10.1098/rsta.2015.0235 }{Von Neumann was not a Quantum
  Bayesian},'' \booktitle{Philosophical Transactions of the Royal
  Society A} \textbf{374} (2016), 20150235, \arxiv{1412.2409}.
    
\bibitem{VanKampen:1988} N.\ G.\ van Kampen,
  ``\hrefdoi{10.1016/0378-4371(88)90105-7}{Ten Theorems about Quantum
  Mechanical Measurements},'' \booktitle{Physica A} \textbf{153}
  (1988), 97--113.

\bibitem{Jaynes:1990} E.\ T.\ Jaynes, ``Probability in quantum
  theory.'' In \booktitle{Complexity, Entropy and the Physics of
    Information} (Addison-Wesley, 1990).

\bibitem{Jaynes:1989} E.\ T.\ Jaynes, ``Cleaning up the mysteries --
  the original goal.'' In \booktitle{Maximum-Entropy and Bayesian
    Methods} (Kluwer, 1989).


\bibitem{Spekkens:2007} R.\ W.\ Spekkens,
  ``\hrefdoi{10.1103/PhysRevA.75.032110}{Evidence for the epistemic
  view of quantum states:~A toy theory},'' \booktitle{Physical Review
  A} \textbf{75} (2007), 032110, \arxiv{quant-ph/0401052}.
    
\bibitem{spekkens-pirsa2016} R.\ W.\ Spekkens, ``Reassessing claims of
  nonclassicality for quantum interference phenomena'' (2016),
  \pirsa{16060102}. 

\bibitem{Pusey:2012} M.\ F.\ Pusey, J.\ Barrett and T.\ Rudolph,
  ``\hrefdoi{10.1038/nphys2309}{On the reality of the quantum
  state},'' \booktitle{Nature Physics} \textbf{8} (2012), 475,
  \arxiv{1111.3328}.
  
\bibitem{qbist-decoherence} C.\ A.\ Fuchs and R.\ Schack, ``Bayesian
  conditioning, the reflection principle, and quantum decoherence,''
  \booktitle{Probability in Physics} (2012), 233--47,
  \arxiv[quant-ph]{1103.5950}.

\bibitem{stacey-thesis} B.\ C.\ Stacey, \booktitle{Multiscale
  Structure in Eco-Evolutionary Dynamics.} PhD thesis, Brandeis
  University, 2015. \arxiv[q-bio.PE]{1509.02958}.

\bibitem{Fuchs:2002} C.\ A.\ Fuchs, ``Quantum mechanics as quantum
  information (and only a little more),'' \arxiv{quant-ph/0205039}
  (2002).  

\bibitem{q-de-finetti1} C.\ M.\ Caves, C.\ A.\ Fuchs and R.\ Schack,
  ``\hrefdoi{10.1063/1.1494475}{Unknown quantum states: the quantum de
  Finetti representation},'' \booktitle{Journal of Mathematical
  Physics} \textbf{43} (2002), 4537--59, \arxiv{quant-ph/0104088}.

\bibitem{q-de-finetti2} C.\ A.\ Fuchs, R.\ Schack and P.\ F.\ Scudo,
  ``\hrefdoi{10.1103/PhysRevA.69.062305}{De Finetti representation
  theorem for quantum-process tomography},'' \booktitle{Physical
  Review A} \textbf{69} (2004), 062305, \arxiv{quant-ph/0307198}.

\bibitem{Barnum:1996} H.\ Barnum, C.\ M.\ Caves, C.\ A.\ Fuchs,
  R.\ Jozsa and B.\ Schumacher,
  ``\hrefdoi{10.1103/PhysRevLett.76.2818}{Noncommuting mixed states
    cannot be broadcast},'' \booktitle{Physical Review Letters}
  \textbf{76} (1996), 2818, \arxiv{quant-ph/9511010}.
  
\bibitem{qs-compat} C.\ M.\ Caves, C.\ A.\ Fuchs and R.\ Schack,
  ``\hrefdoi{10.1103/PhysRevA.66.062111}{Conditions for compatibility
  of quantum-state assignments},'' \booktitle{Physical Review A}
  \textbf{66} (2002), 062111, \arxiv{quant-ph/0206110}.
  
\bibitem{Fuchs:1996} C.\ A.\ Fuchs and A.\ Peres, ``Quantum state disturbance vs.\ information gain: Uncertainty relations for quantum information,'' \booktitle{Physical Review A} \textbf{53} (1996), 2038, \arxiv{quant-ph/9512023}.

\bibitem{Fuchs:1997} C.\ A.\ Fuchs, N.\ Gisin, R.\ B.\ Griffiths,
  C.-S.\ Niu and A.\ Peres, ``Optimal eavesdropping in quantum
  cryptography. I,'' \arxiv{quant-ph/9701039} (1997).

\bibitem{Furusawa:1998} A.\ Furusawa, J.\ L.\ S{\o}rensen,
  S.\ L.\ Braunstein, C.\ A.\ Fuchs, H.\ J.\ Kimble and
  E.\ S.\ Polzik,
  ``\hrefdoi{10.1126/science.282.5389.706}{Unconditional quantum
    teleportation},'' \booktitle{Science} \textbf{282} (1998), 706--9.

\bibitem{Fuchs:2001} C.\ A.\ Fuchs and K.\ Jacobs,
  ``\hrefdoi{10.1103/PhysRevA.63.062305}{Information tradeoff
  relations for finite-strength quantum measurements},''
  \booktitle{Physical Review A} \textbf{63} (2001), 062305,
  \arxiv{quant-ph/0009101}.
  
\bibitem{Peres:2004} A.\ Peres and D.\ Terno,
  ``\hrefdoi{10.1103/RevModPhys.76.93}{Quantum information and
  relativity theory},'' \booktitle{Reviews of Modern Physics}
  \textbf{76} (2004), 93, \arxiv{quant-ph/0212023}.

\bibitem{Fuchs:2017} C.\ A.\ Fuchs, M.\ C.\ Hoang and B.\ C.\ Stacey,
  ``\hrefdoi{10.3390/axioms6030021}{The SIC question: History and
  state of play},'' \booktitle{Axioms} \textbf{6} (2017), 21,
  \arxiv{1703.07901}.

\bibitem{Stacey:2016c} B.\ C.\ Stacey,
  ``\hrefdoi{10.3390/math4020036}{SIC-POVMs and compatibility among
  quantum states},'' \booktitle{Mathematics} \textbf{4} (2016), 36,
  \arxiv{1404.3774}.
  
\bibitem{qplex} M.\ Appleby, C.\ A.\ Fuchs, B.\ C.\ Stacey and H.\ Zhu,
  ``\hrefdoi{10.1140/epjd/e2017-80024-y}{Introducing the Qplex:\ A
  Novel Arena for Quantum Theory},'' \booktitle{The European Physical
  Journal D} \textbf{71} (2017), 197, \arxiv{1612.03234}.

\bibitem{Stacey:2018} B.\ C.\ Stacey, ``Is the SIC outcome there when
  nobody looks?'' \arxiv{1807.07194} (2018).

\bibitem{DeBrota:2017} J.\ B.\ DeBrota and C.\ A.\ Fuchs,
  ``\hrefdoi{10.1007/s10701-017-0098-z}{Negativity bounds for
  Weyl--Heisenberg quasiprobability representations},''
  \booktitle{Foundations of Physics} \textbf{47} (2017), 1009--30,
  \arxiv{1703.08272}.
  
\bibitem{Appleby:2017} M.\ Appleby, S.\ Flammia, G.\ McConnell and
  J.\ Yard, ``\hrefdoi{10.1007/s10701-017-0090-7}{SICs and algebraic
    number theory},'' \booktitle{Foundations of Physics} \textbf{47}
  (2017), 1042--59, \arxiv{1701.05200}.

\bibitem{Kopp:2018} G.\ S.\ Kopp, ``SIC-POVMs and the Stark
  conjectures,'' \arxiv{1807.05877}\\ (2018).

\bibitem{Baez:2012} J.\ C.\ Baez and J.\ Biamonte, ``Quantum
  techniques for stochastic mechanics,'' \arxiv{1209.3632} (2012).

\bibitem{Schlosshauer:2014} M.\ Schlosshauer, ``The
  quantum-to-classical transition and decoherence,'' \arxiv{1404.2635}
  (2014).

\bibitem{Rovelli:1996} C.\ Rovelli,
  ``\hrefdoi{10.1007/BF02302261}{Relational quantum mechanics},''
  \booktitle{International Journal of Theoretical Physics} \textbf{35}
  (1996), 1637--78, \arxiv{quant-ph/9609002}.
    
\bibitem{Stanford} F.\ Laudisa and C.\ Rovelli,
  ``Relational Quantum Mechanics,'' {\sl The Stanford Encyclopedia of
  Philosophy} (2013), Edward N.\ Zalta (ed.),
  \url{http://plato.stanford.edu/entries/qm-relational/}. 

\bibitem{Appleby:2005a} M.\ Appleby,
  ``\hrefdoi{10.1007/s10701-004-2014-6}{Facts, values and quanta},''
  \booktitle{Foundations of Physics} \textbf{35} (2005), 627,
  \arxiv{quant-ph/0402015}.

\bibitem{Appleby:2005b} M.\ Appleby,
  ``\hrefdoi{10.1134/1.2055942}{Probabilities are single-case, or
  nothing},'' \booktitle{Optics and Spectroscopy} \textbf{99} (2005),
  447, \arxiv{quant-ph/0408058}.

\bibitem{SmerlakRovelli2007} M.\ Smerlak and C.\ Rovelli, ``Relational
  EPR,'' \booktitle{Foundations of Physics}\ {\bf 37} (2007), 427--45,
  \arxiv{quant-ph/0604064}.

\bibitem{Ruyant2018} Q.\ Ruyant,
  ``\hrefdoi{10.1007/s10701-018-0156-1}{Can we make sense of
  Relational Quantum Mechanics?}''\ \booktitle{Foundations of Physics}
  \textbf{48} (2018), 440--55.

\bibitem{Voldemort} C.\ A.\ Fuchs, ``QBism, the Perimeter of Quantum
  Bayesianism,'' \arxiv{1003.5209} (2010).
  
\bibitem{Fuchs:2013b} C.\ A.\ Fuchs and R.\ Schack,
  ``\hrefdoi{10.1103/RevModPhys.85.1693}{Quantum-Bayesian
  coherence},'' \booktitle{Reviews of Modern Physics} \textbf{85}
  (2013), 1693--1715, \arxiv{1301.3274}.
  
\bibitem{Mermin:1981} N.\ D.\ Mermin,
  ``\href{https://www.jstor.org/stable/2026482}{Quantum mysteries for
  anyone},'' \booktitle{The Journal of Philosophy} \textbf{78} (1981),
  397--408.

\bibitem{tHooft:1993} G.\ 't Hooft, ``Dimensional reduction in quantum
  gravity,'' \arxiv{gr-qc/9310026} (1993).
  
\bibitem{Witten:2001} E.\ Witten, ``Quantum gravity in de Sitter
  space,'' \arxiv{hep-th/0106109}\\ (2001).
  
\bibitem{Parikh:2005} M.\ K.\ Parikh and E.\ P. Verlinde,
  ``\hrefdoi{10.1088/1126-6708/2005/01/054}{De Sitter holography with
  a finite number of states},'' \booktitle{Journal of High-Energy
  Physics} \textbf{2005} (2005), 54, \arxiv{hep-th/0410227}.
  
\bibitem{Banks:2018} T.\ Banks and W.\ Fischler, ``The Holographic
  Space-Time model of cosmology,'' \arxiv{1806.01749} (2018).

\bibitem{Fuchs:2010} C.\ A.\ Fuchs, ``QBism: The perimeter of Quantum
  Bayesianism,''\\ \arxiv{1003.5209} (2010).
  
\bibitem{Fuchs:2004} C.\ A.\ Fuchs, ``On the quantumness of a Hilbert
  space,'' \booktitle{Quantum Information and Computation} \textbf{4}
  (2004), 467--78, \arxiv{quant-ph/0404122}.

\bibitem{Fuchs:2014} C.\ A.\ Fuchs. \booktitle{My Struggles with the
  Block Universe} (2014). Edited by B.\ C.\ Stacey, with a foreword by
  M.\ Schlosshauer. \arxiv{1405.2390}.
  
\bibitem{Diaconis:2018} P.\ Diaconis and B.\ Skyrms, \booktitle{Ten
  Great Ideas about Chance} (Princeton University Press, 2018).
  
\bibitem{priors} C.\ A.\ Fuchs and R.\ Schack,
  ``\hrefdoi{10.1063/1.3109948}{Priors in quantum Bayesian
  inference}.'' In \booktitle{AIP Conference Proceedings 1101:
  Foundations of Probability and Physics 5}, L. Accardi \emph{et al.,}
  eds. (American Institute of Physics, 2009.)

\bibitem{Savage:1954} L.\ J.\ Savage, \booktitle{The Foundations of
  Statistics} (Dover, 1954).
  
\bibitem{Baez:2016} J.\ C.\ Baez, ``Surprises in logic'' (4 April
  2016). \url{http://math.ucr.edu/home/baez/surprises.html}.
    
  \bibitem{Shulman:2016} M.\ Shulman, ``Homotopy type theory: A
    synthetic approach to higher equalities.'' In
    \booktitle{Categories for the Working Philosopher}, E.\ Landry,
    ed. (Oxford University Press, 2017.)  \arxiv{1601.05035}.
  
\bibitem{Weatherall2011} J.\ O.\ Weatherall,
  ``\hrefdoi{10.1086/660737}{On (some) explanations in physics},''
  \booktitle{Philosophy of Science} \textbf{78} (2011), 421--47.

\bibitem{Distler:2010} J.\ Distler and S.\ Garibaldi, ``\hrefdoi{10.1007/s00220-010-1006-y}{There is no `theory of everything' inside $E_8$},'' \booktitle{Communications in Mathematical Physics} \textbf{298} (2010), 419--36, \arxiv{0905.2658}.
  
\bibitem{Stacey:2017} B.\ C.\ Stacey,
  ``\hrefdoi{10.1007/s10701-017-0087-2}{Sporadic SICs and the normed
  division algebras},'' \booktitle{Foundations of Physics} \textbf{47}
  (2017), 1060--64, \arxiv{1605.01426}.

\bibitem{Stacey:2016} B.\ C.\ Stacey, ``Geometric and
  information-theoretic properties of the Hoggar lines,''
  \arxiv{1609.03075} (2016).

\bibitem{Bouie:2018} J.\ Bouie, ``Taking the Enlightenment seriously
  requires talking about race,'' \booktitle{Slate} (5 June 2018). See
  \url{https://slate.com/news-and-politics/2018/06/taking-the-enlightenment-seriously-requires-talking-about-race.html}.

\bibitem{Switek:2019} B.\ Switek, \booktitle{Skeleton Keys:\ The
  Secret Life of Bone} (Riverhead Books, 2019).
  
\bibitem{ZinnJustin:2007} J.\ Zinn-Justin, \booktitle{Phase
  Transitions and Renormalization Group} (Oxford University Press,
  2007).
  
\bibitem{nlab:2017} nLab authors, ``Lagrangian quantum field theory,'' (27 September 2017). \url{https://ncatlab.org/nlab/show/Lagrangian+quantum+field+theory}.

\bibitem{Baez:2016b} J.\ C.\ Baez, ``Struggles with the Continuum,''
  \arxiv{1609.01421} (2016).

\bibitem{Zee:2010} A.\ Zee, \booktitle{Quantum Field Theory in a
  Nutshell} (Princeton University Press, 2010).
  
\bibitem{Baez:2013} J.\ C.\ Baez and B.\ S.\ Pollard,
  ``\hrefdoi{10.3390/e17020772}{Quantropy},'' \booktitle{Entropy}
  \textbf{17,} 2 (2015), 772--89, \arxiv{1311.0813}.

\bibitem{Pusey:2018} M.\ Pusey, ``\hrefdoi{10.1038/s41567-018-0293-7}{An inconsistent friend},'' \booktitle{Nature Physics} \textbf{14,} 10 (2018), 977--978.

\bibitem{Bub:2018} J.\ Bub, ``\hrefdoi{10.1016/j.shpsb.2018.03.002}{In Defense of a ``Single-World'' Interpretation of Quantum Mechanics},'' \booktitle{Studies in History and Philosophy of Modern Physics} (2018).

\bibitem{workshop:2019} ``Encapsulated Agents in Quantum
  Theory:\ Re-examining Wigner's Friend,''
  \url{https://wigner.mattleifer.info/} (2019).
  
\bibitem{Kent:2015} A.\ Kent,
  ``\hrefdoi{10.1007/s10701-014-9862-5}{Does it make sense to speak of
  self-locating uncertainty in the universal wave function?  Remarks
  on Sebens and Carroll},'' \booktitle{Foundations of Physics}
  \textbf{45,} 2 (2015), 211--17, \arxiv{1408.1944}.
  
  

  


    
\bibitem{Ellis:2017} G.\ F.\ R.\ Ellis, ``Physics on edge''
  (2017). At \url{http://inference-review.com/article/physics-on-edge}.

\bibitem{Wang:2017} Q.\ Wang, Z.\ Zhu and W.\ Unruh,
  ``\hrefdoi{10.1103/PhysRevD.95.103504}{How the huge energy of
  quantum vacuum gravitates to drive the slow accelerating expansion
  of the universe},'' \booktitle{Physical Review D} \textbf{95}
  (2017), 504, \arxiv{1703.00543}.


\bibitem{Nadathur:2014} S.\ Nadathur, ``Does the multiverse explain the cosmological constant?'' Blog post, 3 February 2014. \url{https://blankonthemap.blogspot.com/2014/02/does-multiverse-explain-cosmological.html}.
  
\bibitem{Oesch:2016} P.\ A.\ Oesch \emph{et al.}, ``\hrefdoi{10.3847/0004-637X/819/2/129}{A remarkably luminous galaxy at $z = 11.1$ measured with Hubble Space Telescope grism spectroscopy},'' \booktitle{The Astrophysical Journal} \textbf{819} (2016), 129, \arxiv{1603.00461}.

\bibitem{Weinberg:1987} S. Weinberg, ``\hrefdoi{10.1103/PhysRevLett.59.2607}{Anthropic bound on the cosmological constant},'' \booktitle{Physical Review Letters}\ \textbf{59} (1987), 2607.
  
\bibitem{Mermin-Bell} N.\ D.\ Mermin, ``Why QBism is not the
  Copenhagen interpretation and what John Bell might have thought of
  it,'' \arxiv[quant-ph]{1409.2454}.  In \booktitle{Quantum
    [Un]Speakables II} (Springer-Verlag, 2017).

\bibitem{Fuchs:2012} C.\ A.\ Fuchs, ``Interview with a Quantum
  Bayesian,'' \arxiv{1207.2141} (2012). This collects the
  contributions by Fuchs to M.\ Schlosshauer's \booktitle{Elegance and
    Enigma: The Quantum Interviews} (Springer, Frontiers Collection,
  2011).
  
\bibitem{Chalmers2014} M.\ Chalmers, ``QBism: Is
  quantum uncertainty all in the mind?''\ \booktitle{New Scientist}
  issue 2968 (2014): 32--25.

\bibitem{VonBaeyer2013} H.\ C.\ von Baeyer, ``Can Quantum Bayesianism
  Fix the Paradoxes of Quantum Mechanics?''\ \booktitle{Scientific
    American} {\bf 308} (2013):
  46--51.

\bibitem{Mermin14b} N.\ D.\ Mermin, ``QBism in the New Scientist,''
  \arxiv[quant-ph]{1406.1573}.

\bibitem{Mermin-Vienna} N.\ D.\ Mermin, ``Putting the Scientist into
  the Science,'' lecture at \booktitle{Quantum [Un]Speakables II: 50
    Years of Bell's Theorem,} University of Vienna (2014).
  \url{https://phaidra.univie.ac.at/detail_object/o:360625}.

\bibitem{VonRauchhaupt2014} U.\ von Rauchhaupt,
  ``\href{http://www.faz.net/aktuell/wissen/physik-chemie/philosophische-quantenphysik-ganz-im-auge-des-betrachters-12792104.html}{Philosophische
  Quantenphysik : Ganz im Auge des Betrachters}'' [Philosophical
  Quantum Physics: Only in the Eye of the Beholder],
  \booktitle{Frankfurter Allgemeine Sonn\-tagszeitung} (9 February 2014).

\bibitem{Ball2013} P.\ Ball,
  ``\hrefdoi{10.1038/501154a}{Quantum Quest},''
  \booktitle{Nature} {\bf 501} (2013): 154--56.

\bibitem{philzone} J.\ Gelonasi and C.\ A.\ Fuchs, ``Quantum Worlds,''
  \booktitle{The Philosopher's Zone}
  (2015). \url{http://www.abc.net.au/radionational/programs/philosopherszone/}.

\end{thebibliography}
\end{document}